\documentclass[12pt]{iopart}
\usepackage{iopams}
\usepackage{psfrag,graphicx}
\usepackage{graphicx}
\usepackage{subfigure}
\usepackage[section]{placeins}
\usepackage{psfrag}
\usepackage{caption2}

\begin{document}
\title[Coupling characteristics of AS$_2$S$_3$ chalcogenide PCF coupler]{A projection operator approach for computing the dynamics of AS$_2$S$_3$ chalcogenide birefringent photonic crystal fiber coupler}
\author{T. Uthayakumar$^1$, R. Vasantha Jayakantha Raja$^{*2}$ and K. Porsezian$^1$}
\address{${^1}$Department of Physics, Pondicherry University, Puducherry - 605014, India.\\
$^2$Centre for Nonlinear Science and Engineering, School of Electrical and Electronics Engineering, SASTRA University, Thanjavur 613401, India.}
\ead{uthayapu@gmail.com, rvjraja@yahoo.com and ponzsol@yahoo.com}

\begin{abstract}
A variety of AS$_2$S$_3$ chalcogenide photonic crystal fiber coupler (CPCFC) of special properties are proposed to study the role of birefringence in all optical coupling characteristics based on projection operator method (POM). The equations of motion describing the dynamics of the individual pulse parameters through x- and y- polarized modes are arrived by employing POM from the coupled nonlinear Schr\"{o}dinger equations (CNLSE). From the pulse parameter dynamics, it is observed that the amplitude of the polarization components are significantly influenced by the pulse introduced with different polarizing angle even at low input power level. Such, selective polarizing angle  of the input pulse will provide the efficient control over the desired splitting ratio as well as the ability to decide the desired polarization component.
\end{abstract}

\maketitle

\section{Introduction}
\begin{figure}[tbp]
\begin{center}
\label{pcfcoupler}
\includegraphics[width=0.6\textwidth]{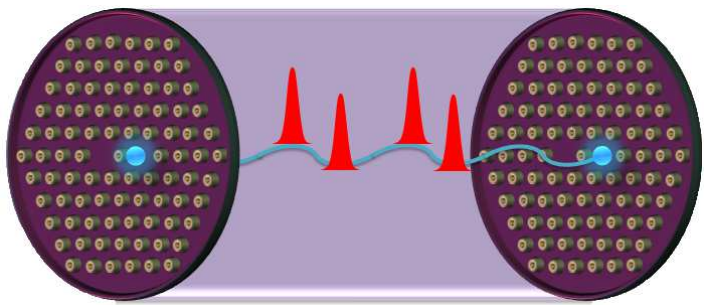}
\caption{Schematic diagram of PCF coupler.}
\end{center}
\end{figure}

All optical couplers play a vital role in the conventional communication networks as couplers, splitters, switches, multiplexors and optical controllers \cite{Agrawal}. Such devices fabricated from photonic crystal fiber (PCF) attracted considerable interest for their unusual light guiding properties which cannot be afforded by conventional fiber couplers. Dual core couplers constructed from PCF exhibit many exciting optical properties namely enhanced nonlinearity, high dispersion, high birefringence, desired zero dispersion wavelengths, low confinement loss, etc \cite{Saitoh1,Gerome,Fogli,Zhang,Uthayakumar1,Uthayakumar2,Saitoh2}. A schematic of the dual core PCF couplers (PCFC) is shown in Fig. \ref{pcfcoupler}. Such PCF coupler, fabricated from PCF exhibits the merit of easy fabrication and provides the possibility of achieving the birefringence of many orders of magnitude compared to that of the conventional optical fibers. Numerous contributions have been made both theoretically as well as experimentally for achieving the efficient birefringence using PCF for a variety of applications. The impact of fiber and pulse parameters on birefringent PCF to evaluate the optical properties namely mode field diameter, modal birefringence, divergence angle and polarization mode dispersion for sensing and photonic applications have been already investigated \cite{Gerosa,Ortigosa,Hansen2,Ju,Chen,Liou,Saitoh3}. Studies on selective coupling in high polarization sensitivity couplers also made for their applications as mulitplexors and optical interconnects by employing microstructures \cite{Gerosa}.

In recent years, the optical fibers and other devices fabricated from non-silica glasses have attracted for their high nonlinearity contribution at infrared region \cite{Anton,Monro}. In particular, chalcogenide glasses received greater attention for their high refractive, enhanced nonlinearity of many orders of magnitude compared to the silica. The optical guiding and dispersion characteristics of the chalcogenide PCF in the infrared wavelength region using hexagonal and square lattice chalcogenide glass have been explored successfully by employing improved fully vectorial effective index method \cite{Dabas}. A variety of highly nonlinear PCFs with dispersion optimization have also been successfully designed using finite element method (FEM) for various all optical applications at 1.55 $\mu$m \cite{Kanka}. The switching characteristics of the highly nonlinear chalcogenide PCFC (CPCFC) was first successfully demonstrated numerically by combining coupled mode theory and fully-vectorial integral equation analysis \cite{Chremmos}. The dynamics of the individual pulse parameter dynamics and switching characteristics of the twin core highly nonlinear CPCFC has been successfully demonstrated by split step Fourier method (SSFM) \cite{Uthayakumar3}.

The coupling characteristics and polarization properties of chalcogenide coupler for mid-infrared wavelength have been explored by employing asymmetric cores \cite{Shuo}. A detailed review of linear and nonlinear properties of chalcogenide glasses for all optical applications are discussed by Zakery et al \cite{Zakery}. The switching characteristics of the highly birefringent nonlinear two core fiber couplers are dealt with varying polarization angle to unravel the influence of group delay dispersion and coupling coefficient dispersion (CCD) \cite{Li2}. But till date, to the best of our knowledge, there is no significant contribution to the study of the individual pulse parameter dynamics and influence of polarization angle of highly nonlinear birefringent PCFC at 1.55 $\mu$m.  However, the PCFC formed in a highly nonlinear birefringent chalcogenide forms the best candidate for achieving efficient all optical control and sensing devices. Hence, we are interested in investigating the pulse dynamics to reveal physical insight of the individual pulse parameters during the propagation, coupling and energy swap analytically.  There are many analytical techniques like the Lagrangian variational method \cite{Anderson}, Hamiltonian method \cite{Kutz}, projection operator method (POM) \cite{Nakkeeran1}, non-Lagrangian collective variable approach \cite{Moubissi}, collective variable technique \cite{dindacv} and the moment method \cite{Agrawal} available to study the dynamics governed by CNLSE.  Among these analytical techniques, we choose the generalized POM to investigate the switching characteristics of nonlinear coupler because, the POM is considered to be more versatile as it does not require the complex procedure of derivation of the Lagrangian.

In a previous report, we already introduced the concept of POM technique in PCF coupler to investigate the switching dynamics in detail \cite{Uthayakumar1}. Our further investigation has revealed the strange role of pulse dynamics in birefringent PCF coupler. Thus, in this paper, we intend to study the dynamics of the highly nonlinear birefringent PCFC by appropriate CPCFC structure using POM. Furthermore, we are also inquisitive to explore the consequence of varying air hole radius along the slow axis on the light guiding characteristics. Hence, we put forth three highly birefringent CPCFC with varying air hole radius for our study to construct it much closer to that of the real systems. The study is outlined as follows: The design of the highly nonlinear CPCFC and the calculation of the optical parameters are provided in section 2. A theoretical study of the individual pulse parameter dynamics described by CNLSE obtained through the projection operator method (POM) is discussed by section 3. The section 4 deals with the coupling characteristics of the proposed geometries. Finally, the section 5 concludes the paper.

\section{Modeling of highly nonlinear birefringent CPCFC}
\begin{figure}[tbp]
\label{pcfc}
\includegraphics[width=0.5\textwidth]{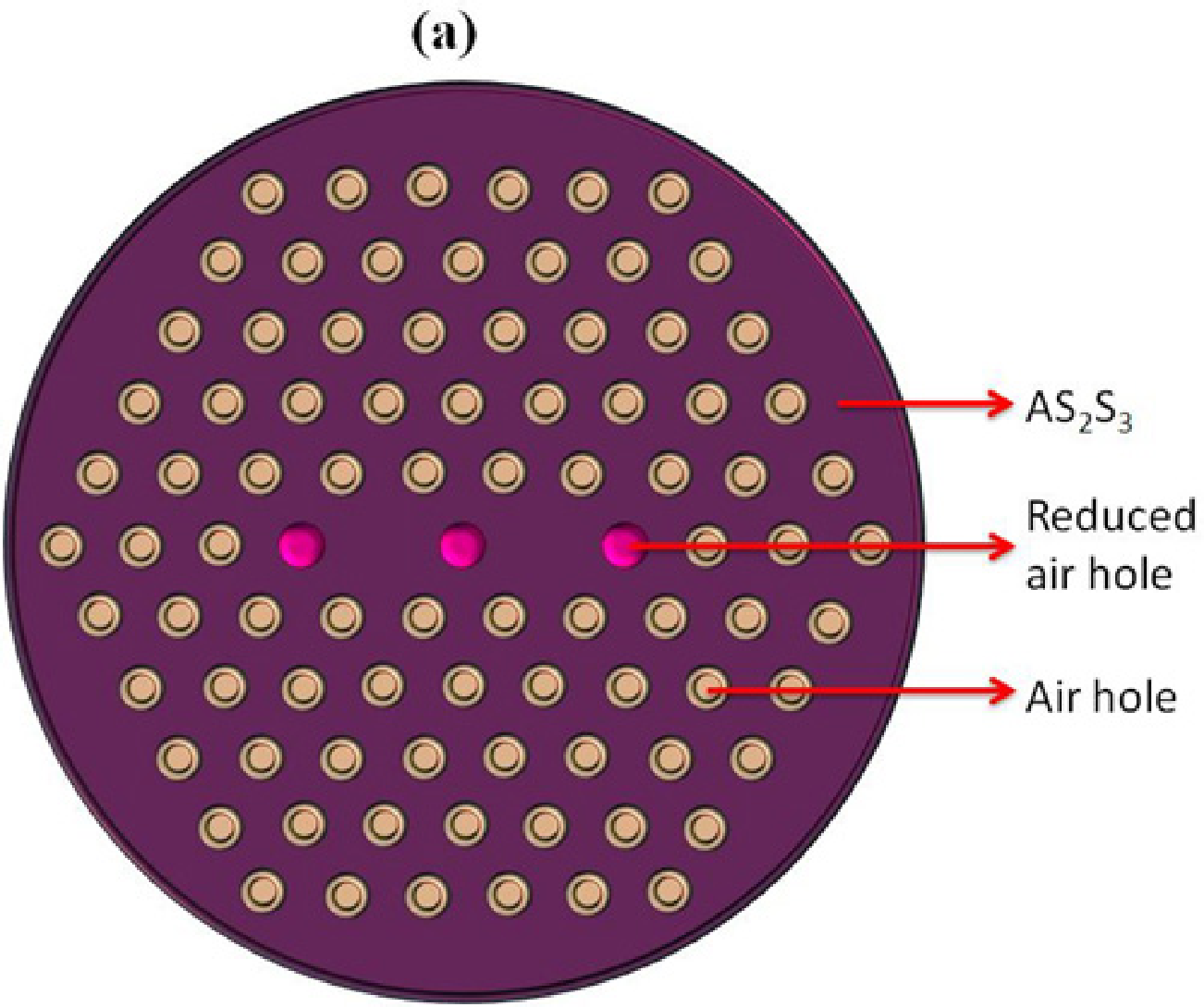}
\label{lambdavsindex}
\includegraphics[width=0.5\textwidth]{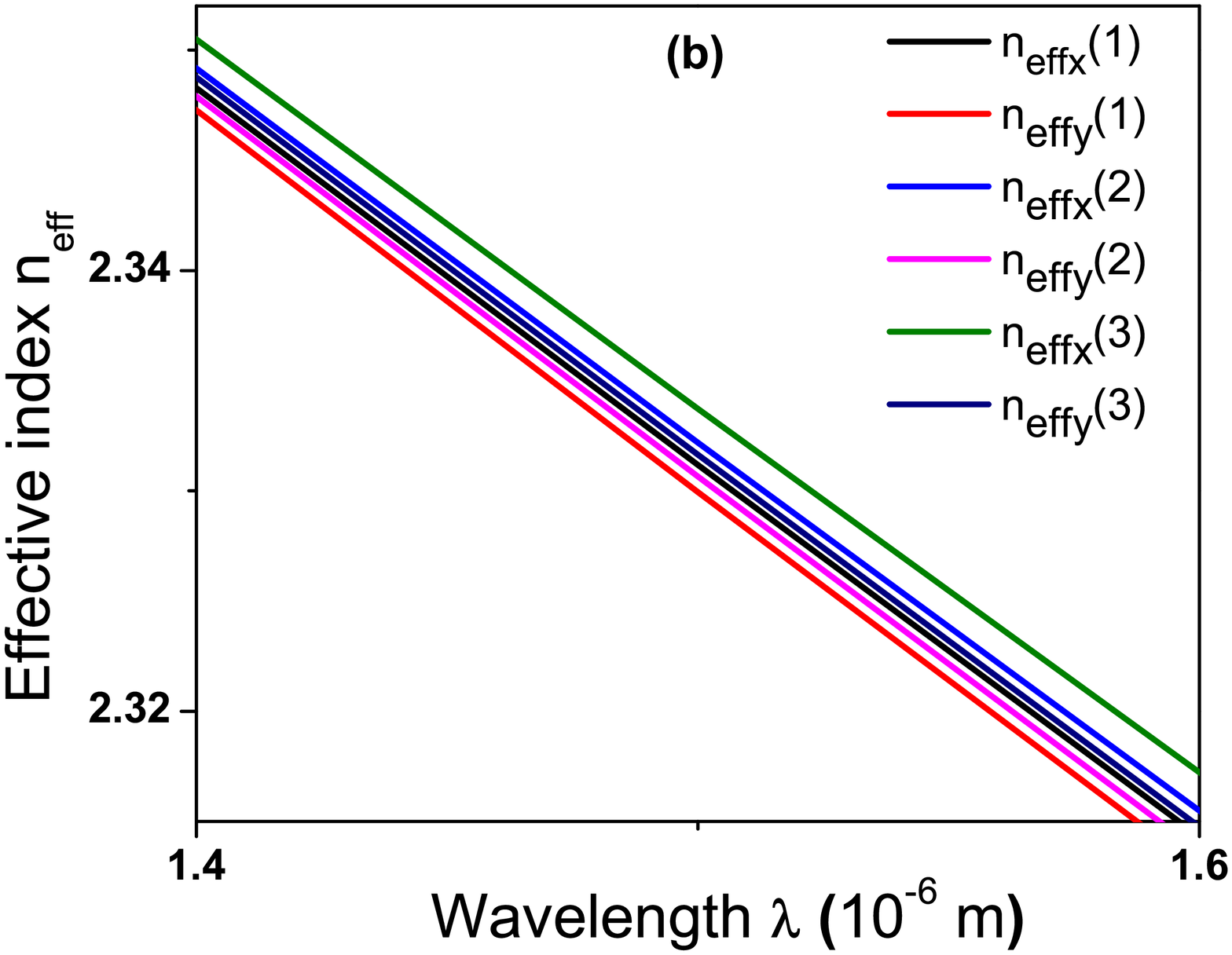}
\caption{(a) Schematic of highly nonlinear birefringent CPCFC with structure parameters d/$\Lambda$ = 0.7 and $\Lambda$ = 1 $\mu$m and (b) Variation of refractive index as a function of the wavelength. }
\end{figure}

In order to explore the influence of the varying air hole diameter on birefringence in highly nonlinear birefringent coupler, we put forth three types of arsenic trisulfide (AS$_2$S$_3$) CPCFC. The schematic of the proposed CPCFC is shown in Fig. \ref{pcfc} (a). The structure parameters of the proposed CPCFC designs are air hole diameter ($d$) to the pitch constant ($\Lambda$) ratio d/$\Lambda$ = 0.7 and $\Lambda$ = 1 $\mu$m. In order to obtain birefringence, the diameter of the central air hole and two air holes adjacent to the guiding core indicated by pink circle in Fig. \ref{pcfc} (a) are reduced accordingly. For the designs 1, 2 and 3 the diameter of the air holes (pink colored) are reduced 5$\%$, 7$\%$ and 10$\%$ respectively. The x-axis with high effective index becomes slow axis and y-axis with low effective index turns into fast axis for the proposed structures. The effective refractive indices of the even and odd polarized x- and y- modes of CPCFC, are calculated using the FEM \cite{Saitoh3,Uthayakumar4}. For FEM calculations of the proposed geometry, we have used 37504 triangular elements and the boundary conditions are provided by \cite{Saitoh3}. From the calculated effective refractive indices, the numerical values of first order dispersion ($\beta_1$), walk-off length, dispersion, group velocity dispersion parameter ($\beta_2$), nonlinearity, birefringence, beat length, coupling length and CCD of the even and odd symmetry supporting x- and y- polarized modes of all structures are determined. The structure and the optical parameters determined for the CPCFC designs are tabulated in Table. 1.

Fig. \ref{pcfc} (b) shows the dependence of refractive index as a function of the wavelength.  At shorter wavelengths, effective refractive index is maximum and index difference between the birefringent axes also minimum. As the
wavelength increases, the effective index decreases with gradual increase in index difference between the fast and
slow axis. A considerably greater index difference between the birefringent axes is achieved at higher wavelength end.  From the figure, it is clearly observed that structure with 10$\%$ reduction in air hole diameter (design 3) shows considerable index difference between the birefringent axes which is illustrated by green and dark blue lines of Fig. \ref{pcfc} (b).

\begin{table}
\caption{\textbf{The design and the optical parameters of the chalcogenide PCFC}}
\renewcommand{\arraystretch}{1.4}
\addtolength{\tabcolsep}{-7pt}
\begin{tabular}{|l|l|l|l|l|l|l|}
\hline
\multicolumn{7}{|l|}{\,\,\,\,\,\,\,\,\,\,\,\,\,\,\,\,\,\,\,\,\,\,\,\,\,\,\,\,\,\,\,\,\,\,\,\,\,\,\,\,\,\,\,\,\,\,\,\,\,\,\,\,\,\,\,\,\,\,\,\,\,\,\,\,\,\,\,\,\,\,\,\,\,\,\,\,\,\,\,\,\,\,\,\,\,\,\,\,\,\,\textbf{Design Parameters}} \\
\hline
\,\,\,\,\,\,\,\,\,\,\,\,\,\,\textbf{Parameters}&\multicolumn{2}{l|}{\,\,\,\,\,\,\,\,\,\,\,\,\,\,\textbf{Design 1}}&\multicolumn{2}{l|}{\,\,\,\,\,\,\,\,\,\,\,\,\,\,\textbf{Design 2}}&\multicolumn{2}{l|}{\,\,\,\,\,\,\,\,\,\,\,\,\,\,\textbf{Design 3}}\\
\hline
\,\,\,\,\,\,\,\,\,\,\,\,\,\,\,\,\,\,\,\,\,\,\,\,\,d/$\Lambda$&\multicolumn{2}{l|}{\,\,\,\,\,\,\,\,\,\,\,\,\,\,\,\,\,\,\,\,\,\,\,0.7}&\multicolumn{2}{l|}{\,\,\,\,\,\,\,\,\,\,\,\,\,\,\,\,\,\,\,\,\,\,\,0.7}&\multicolumn{2}{l|}{\,\,\,\,\,\,\,\,\,\,\,\,\,\,\,\,\,\,\,\,\,\,\,0.7}\\
\hline
\,\,\,\,\,\,\,\,\,\,\,\,\,\,\,\,\,\,\,\,\,\,\,\,\,\,\,$\Lambda$&\multicolumn{2}{l|}{\,\,\,\,\,\,\,\,\,\,\,\,\,\,\,\,\,\,\, 1\,$\mu$m}&\multicolumn{2}{l|}{\,\,\,\,\,\,\,\,\,\,\,\,\,\,\,\,\,\,\, 1\,$\mu$m}&\multicolumn{2}{l|}{\,\,\,\,\,\,\,\,\,\,\,\,\,\,\,\,\,\,\,\,1\,$\mu$m}\\
\hline
\,\,\,\,\,\,\,\,\,\,\,\,diameter (d) of &\multicolumn{2}{l|}{
\,\,\,\,\,\,\,\,\,\,\,5\% reduction}&\multicolumn{2}{l|}{
\,\,\,\,\,\,\,\,\,\,\,7\% reduction}&\multicolumn{2}{l|}{
\,\,\,\,\,\,\,\,\,10\% reduction}\\
\,\,\,\,the air holes 1, 2 $\&$ 3&\multicolumn{2}{l|}{
\,\,\,\,\,\,\,\,\,\,\,\,\,\,in diameter}&\multicolumn{2}{l|}{
\,\,\,\,\,\,\,\,\,\,\,\,\,\,in diameter}&\multicolumn{2}{l|}{
\,\,\,\,\,\,\,\,\,\,\,\,\,\,in diameter}\\
\hline
\multicolumn{7}{|l|}{\,\,\,\,\,\,\,\,\,\,\,\,\,\,\,\,\,\,\,\,\,\,\,\,\,\,\,\,\,\,\,\,\,\,\,\,\,\,\,\,\,\,\,\,\,\,\,\,\,\,\,\,\,\,\,\,\,\,\,\,\,\,\,\,\,\,\,\,\,\,\,\,\,\,\,\,\,\,\,\,\,\,\,\,\,\,\,\,\textbf{Optical Parameters}} \\
\hline
\,\,\,\,\,\,\,\,\,\,\,\,\textbf{At 1.55 $\mu$m}&\,\,\,\,\,\,\,\,\textbf{Slow}&\,\,\,\,\,\,\,\textbf{Fast}&\,\,\,\,\,\,\textbf{Slow}&\,\,\,\,\,\,\,\,\textbf{Fast}&\,\,\,\,\,\,\,\,\textbf{Slow}&\,\,\,\,\,\,\,\,\textbf{Fast}\\
\hline
\,\,\,\,\,\,\,\,\,\,\,\,\,\,$\beta_1$\,(ps/m)&\,\,\,\,0.86151&\,\,\,\,0.86218&\,\,\,\,0.86109&\,\,\,\,0.86204&\,\,\,\,0.86046&\,\,\,\,0.86178\\
\hline
\,\,\,\,Walk-off length (m)&\,\,1.21$\times10^3$&\,\,1.39$\times10^3$&\,\,1.12$\times10^3$&\,\,1.29$\times10^3$&\,\,\,1.00$\times10^3\,$&\,\,\,1.15$\times10^3\,$\\
\hline
\,\,\,\,\,\,\,\,\,\,\,\,\,\,$\beta_2$\,(ps$^2$/m)&\,\,-0.08357&\,\,-0.09727&\,\,-0.07593&\,\,-0.09483&\,\,-0.06425&\,\,-0.09052\\
\hline
\,Dispersion (ps/nm/km)\,&\,\,\,65.76539\,\,&\,\,76.54768&\,\,\,59.74943&\,\,74.62814&\,\,\,50.55934&\,\,71.23747\\
\hline
\,Nonlinearity (W$^{-1}$m$^{-1}$)&\,\,\,\,1.01270&\,\,\,\,1.17110&\,\,\,\,1.00300&\,\,\,\,1.31100&\,\,\,\,0.98821&\,\,\,\,1.30582\\
\hline
\,\,\,\,\,\,\,\,\,\,\,\,\,\,\,Birefringence&\multicolumn{2}{l|}{\,\,\,\,\,\,\,\,\,\,\,\,\,\,\, 0.00132}&\multicolumn{2}{l|}{\,\,\,\,\,\,\,\,\,\,\,\,\,\,\,\,0.00168}&\multicolumn{2}{l|}{\,\,\,\,\,\,\,\,\,\,\,\,\,\,\,\, 0.00227}\\
\hline
\,\,\,\,\,\,Beat length (ps/m)&\multicolumn{2}{l|}{\,\,\,\,\,\,\,\,\,\,\,\,\,\,\, 1.17x10$^{-3}$}&\multicolumn{2}{l|}{\,\,\,\,\,\,\,\,\,\,\,\,\,\,\, 9.23x10$^{-4}$}&\multicolumn{2}{l|}{\,\,\,\,\,\,\,\,\,\,\,\,\,\,\,6.84x10$^{-4}$}\\
\hline
\,\,\,\,Coupling length (m)&\,\,\,\,0.00138&\,\,\,\,0.00185&\,\,\,\,0.00125&\,\,\,\,0.00167&\,\,\,\,0.00108&\,\,\,\,0.00143\\
\hline
\,\,\,\,\,\,\,\,\,\,\,\,\,\,\,\,CCD (ps/m)&\,1.37x10$^{-6}$&\,2.17x10$^{-6}\,$&\,1.21x10$^{-6}\,\,$&\,1.88x10$^{-6}\,\,$&\,1.01x10$^{-6\,}$&\,1.57x10$^{-6}\,$\\
\hline
\end{tabular}
\end{table}

\begin{figure}[tbp]
\label{beta1}
\includegraphics[width=0.5\textwidth]{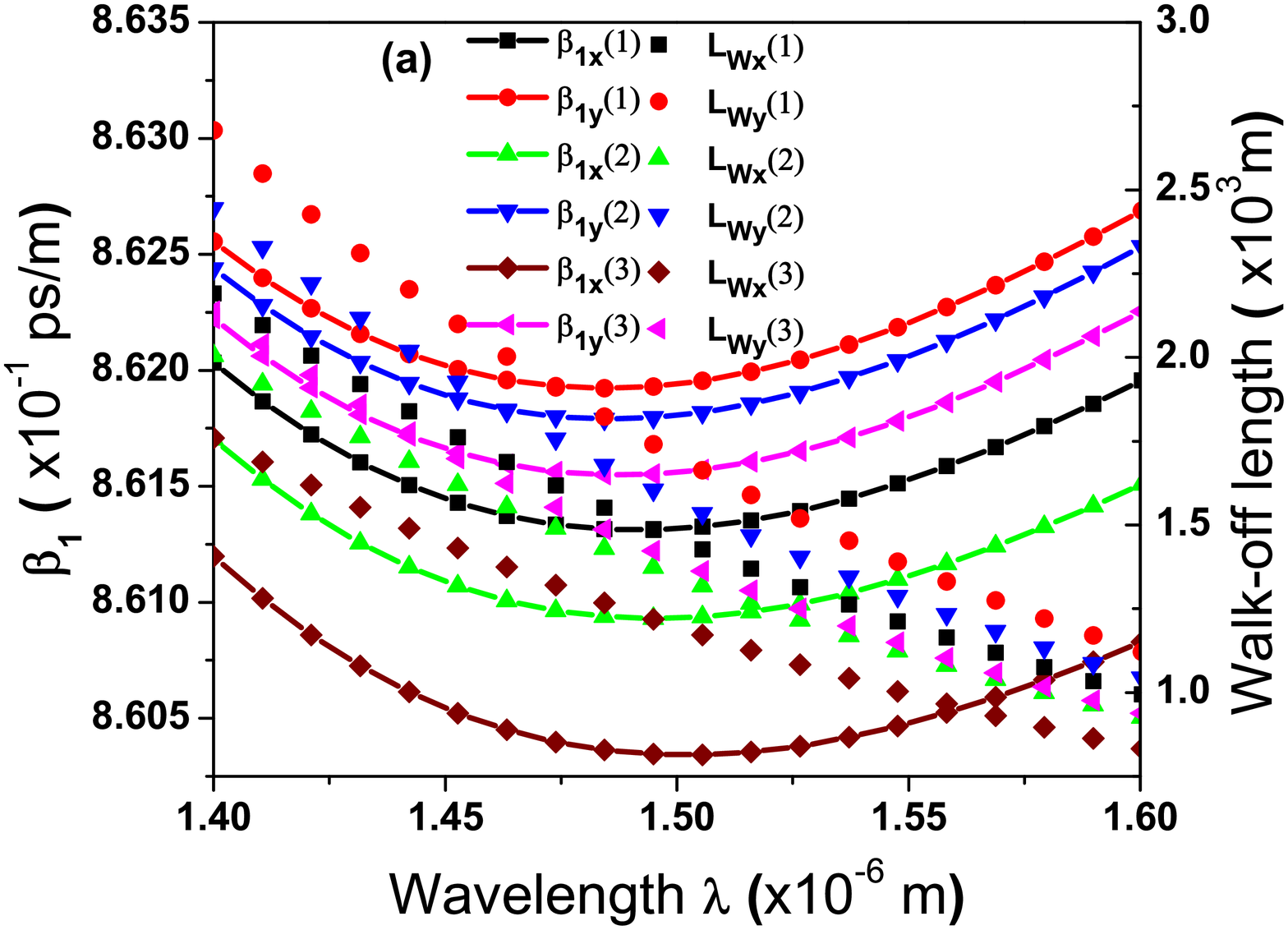}
\label{beta2}
\includegraphics[width=0.5\textwidth]{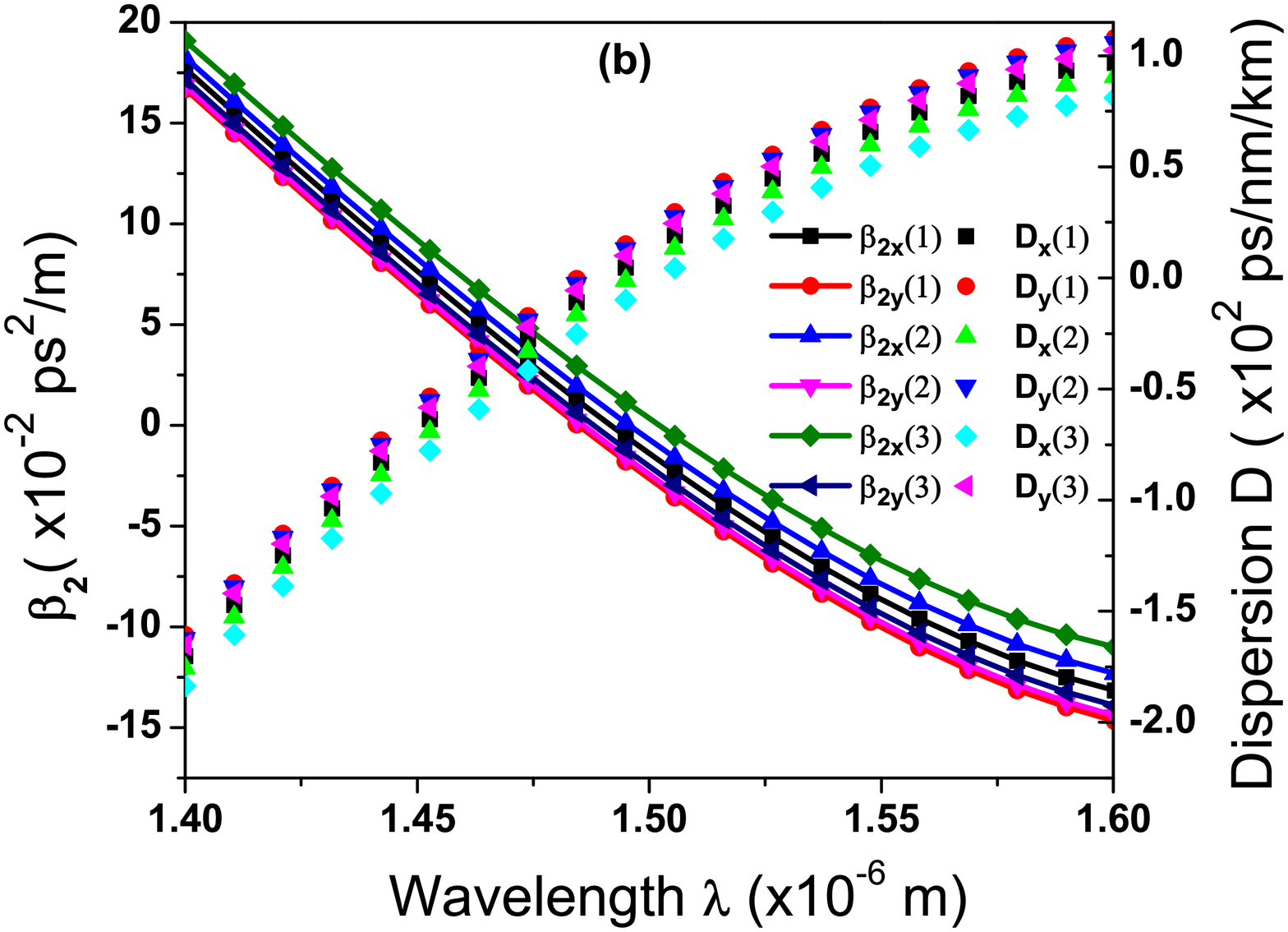}
\caption{Variation of (a) $\beta_1$ and walk-off length and (b) $\beta_2$ and dispersion as a function of the wavelength.}
\end{figure}
Fig. 3 (a) shows the variation of $\beta_1$ and walk-off length of birefringent modes of the PCFC as the function of the wavelength. Variation of $\beta_1$ is represented by the symbol with line and that of walk-off length is portrayed by symbol without line. In all the cases, the $\beta_1$ in the fast axis dominates that of the slow axis.  Initially, the $\beta_1$ decreases gradually with increase in wavelength and it becomes minimum at wavelength around 1.48 $\mu$m. This minimum value of the $\beta_1$ shifts gradually towards 1.50 $\mu$m with the reduction in the diameter of the air holes. And for further increase in wavelength it gradually increases. Moreover, the variation of the $\beta_1$ between the birefringent axes increases with the reduction in diameter of the air hole. From the figure, one can clearly observe that the $\beta_1$ for design 1 is greater than that of the other two designs. Similarly, walk-off length also found to be maximum for fast axis and minimum for the slow axis. Furthermore, the walk-off length gradually reduces as the function of the wavelength for all the cases. And for wavelength around 1.4 $\mu$m walk-off length of birefringent axes show greater variation in their values and decreases with increase in wavelength. Walk-off length is found to be minimum and becomes closer for all the designs at the longer wavelength region.
\begin{figure}[tbp]
\label{birefbeatlength}
\includegraphics[width=0.5\textwidth]{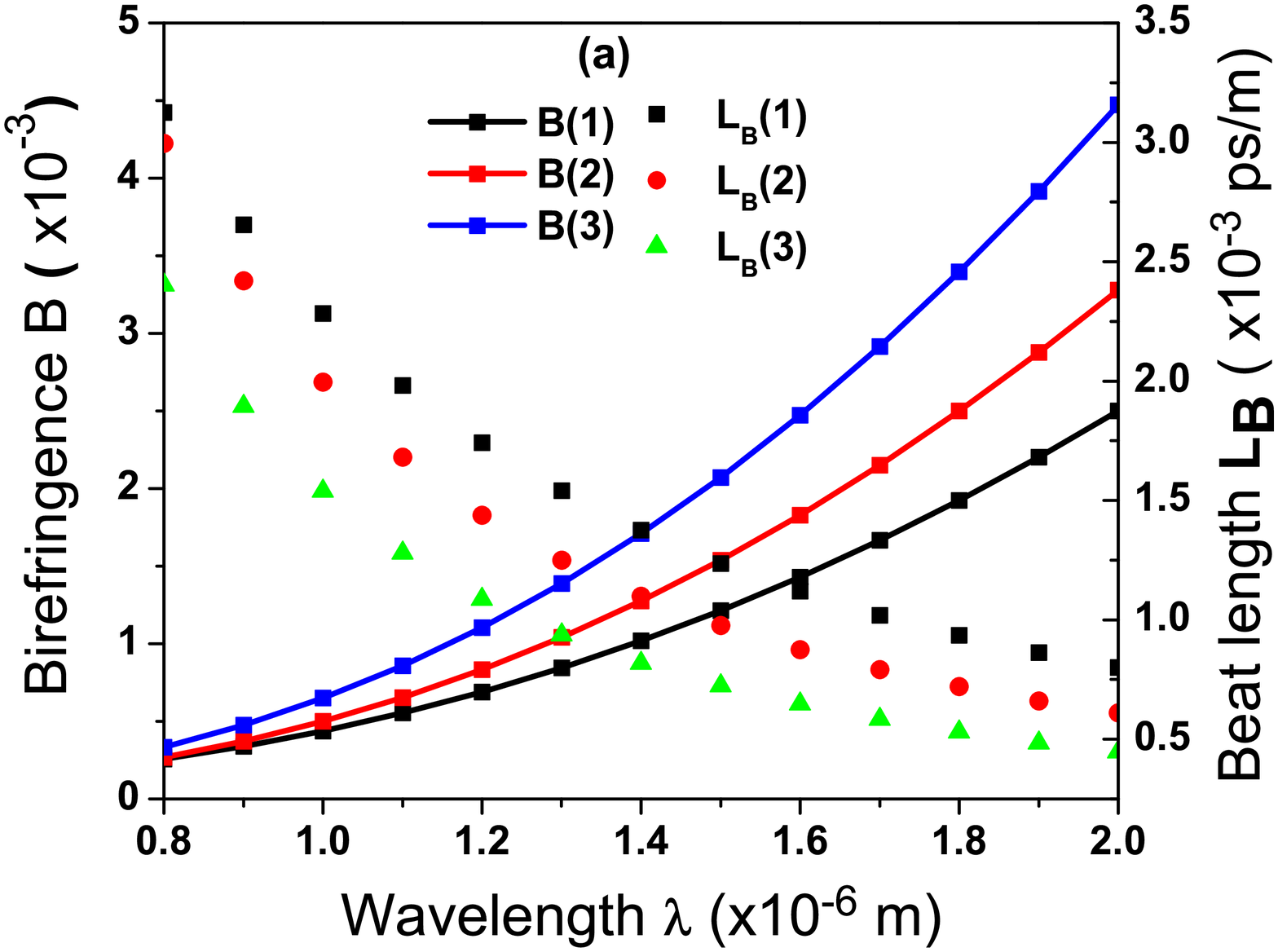}
\label{kappaccd}
\includegraphics[width=0.5\textwidth]{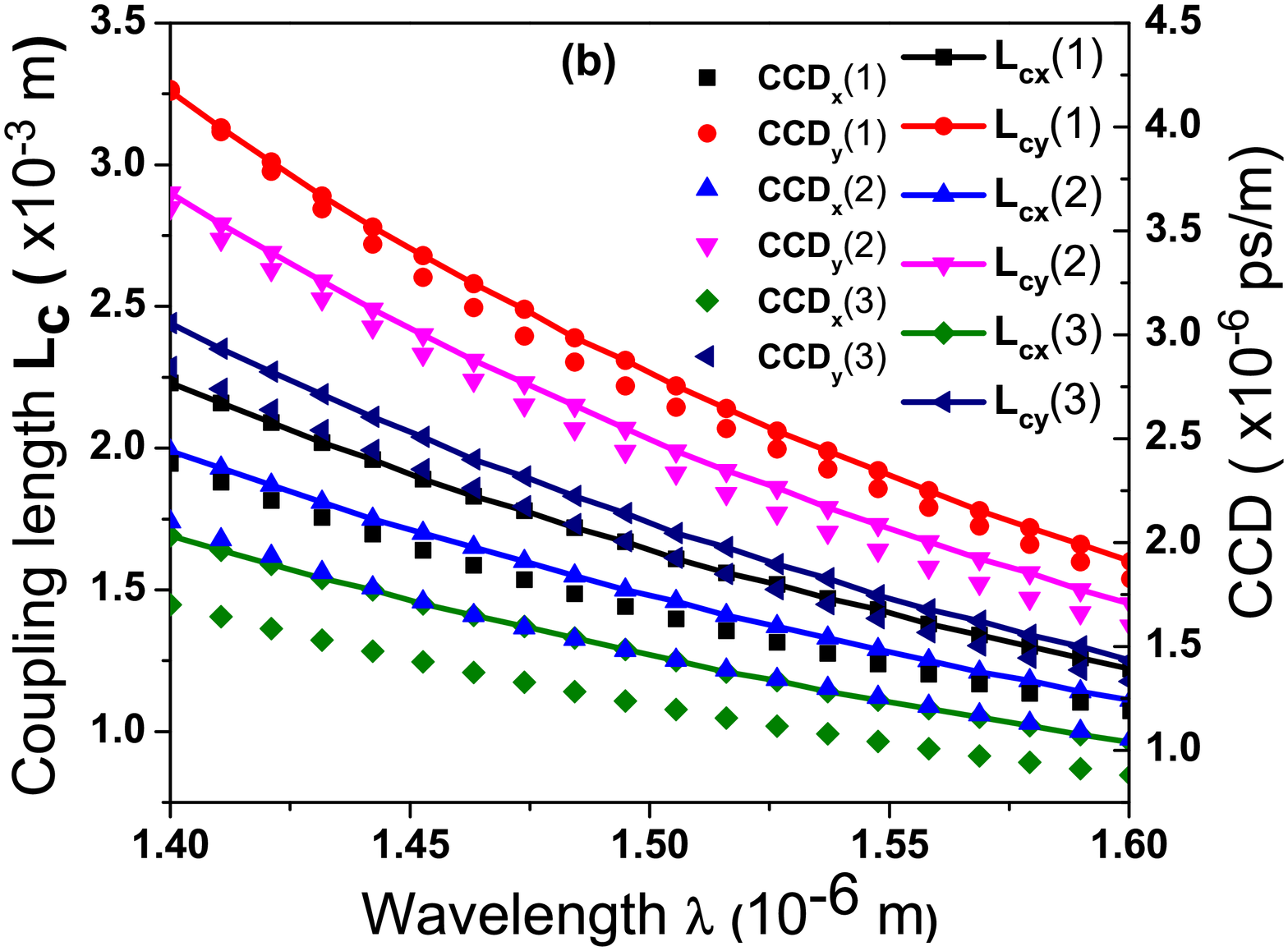}
\caption{Variation of (a) Birefringence and beat length and (b) coupling length and coupling coefficient dispersion as a function of the wavelength.}
\end{figure}

Fig. 3 (b) shows the variation of $\beta_2$ and dispersion with respect to the wavelength. For all the cases, the $\beta_2$ through the x polarized mode is greater than the y polarized mode due to the greater effective index. $\beta_2$ is found to increase with reduction in the diameter of the air holes. Initially, the $\beta_2$ is positive and it gradually decreases with increase in wavelength. And it becomes negative around 1.48 $\mu$m and decreases further.  Whereas the dispersion increases from negative and approaches the positive value with increase in wavelength. Dispersion is negative up to the wavelength around 1.48 $\mu$m and thereafter becomes positive as illustrated in Fig. 3 (b). Dispersion found to decrease with the reduction in the size of the air hole and it is maximum along the slow axis. $\beta_2$ is found to be maximum for design 3 while dispersion is maximum for design 1.

Variation of birefringence and beat length (L$_B$) as the function of wavelength is portrayed in Fig. 4 (a). Birefringence is found to increase gradually with increase in wavelength and maximum birefringence of 2.27x10$^{-3}$ is observed for the design 3 at 1.55 $\mu$m.  L$_B$ is found to decrease with increase in wavelength and design 1 exhibits maximum beat length when compared to that of other designs. Design 3 exhibits the least (L$_B$) of 6.84x10$^{-4}$ ps/m at 1.55 $\mu$m which can preserve polarization for great extent compare to other designs. Thus, the splitting of x- and y- polarized components of the design 3 can be achieved within much shorter distance. The dependence of coupling length (L$_c$) and coupling coefficient dispersion (CCD) is illustrated by Fig. 4 (b).  Both L$_c$ and CCD are found to decrease with increase in wavelength and takes the maximum value for fast axis. The difference between L$_c$ as well as CCD of the birefringent axes decreases as one move from design 1 to design 3. The slow axis of the design 3 exhibits the shortest L$_c$ as well as least CCD.
\section{Theoretical Model}
\label{Theory}
When an optical signal of low power is introduced through any one of the input port of birefringence PCFC, the signal experiences different effective index along the birefringent axes and splits into x- and y- polarized components when it reaches birefringence L$_B$.  When these two components reaches their individual specific distance known as L$_c$ as they propagate through the coupler, they completely transfer to the neighboring core with an introduction of phase shift. And when it reaches a distance L = L$_c$, it again returns to the original core where it is originally launched. For further increase in distance, the input signal oscillates between the cores as it propagates. The dynamics of such birefringent nonlinear couplers is described by a set of coupled nonlinear Schr\"{o}dinger equations (CNLSE) given by \cite{Agrawal,Li2,Li}
\begin{eqnarray}
\label{CNLSE1}i\bigg(\frac{\partial u_x}{\partial z}+\beta_{1x}\frac{\partial u_x}{\partial t}\bigg)-\frac{\beta_{2x}}{2}\frac{\partial^2 u_x}{\partial t^2}+\gamma_x(|u_x|^2+\sigma |u_y|^2)u_x+\kappa_x v_x+i\kappa_{1x} \frac{\partial v_x}{\partial t}\nonumber\\-i\frac{\alpha_x}{2} u_x=0,\\
\label{CNLSE2}i\bigg(\frac{\partial v_x}{\partial z}+\beta_{1x}\frac{\partial v_x}{\partial t}\bigg)-\frac{\beta_{2x}}{2}\frac{\partial^2 v_x}{\partial t^2}+\gamma_x(|v_x|^2+\sigma |v_y|^2)v_x+\kappa_x u_x+i\kappa_{1x} \frac{\partial u_x}{\partial t}\nonumber\\-i\frac{\alpha_x}{2} v_x=0,\\
\label{CNLSE3}i\bigg(\frac{\partial u_y}{\partial z}+\beta_{1y}\frac{\partial u_y}{\partial t}\bigg)-\frac{\beta_{2y}}{2}\frac{\partial^2 u_y}{\partial t^2}+\gamma_y(|u_y|^2+\sigma |u_x|^2)u_y+\kappa_y v_y+i\kappa_{1y} \frac{\partial v_y}{\partial t}\nonumber\\-i\frac{\alpha_y}{2} u_y=0,\\
\label{CNLSE4}i\bigg(\frac{\partial v_y}{\partial z}+\beta_{1y}\frac{\partial v_y}{\partial t}\bigg)-\frac{\beta_{2y}}{2}\frac{\partial^2 v_y}{\partial t^2}+\gamma_y(|v_y|^2+\sigma |v_x|^2)v_y+\kappa_y u_y+i\kappa_{1y} \frac{\partial u_y}{\partial t}\nonumber\\-i\frac{\alpha_y}{2} v_y=0.
\end{eqnarray}
where $u_j$ and $v_j$ with j = x, y are the j$^{th}$ polarized amplitudes of the core 1 and core 2 respectively. $\beta_{1j}$ and $\beta_{2j}$ are group delay and group velocity dispersion of the respective j$^{th}$ polarization respectively. $\gamma_j$ is nonlinearity parameter of the j$^{th}$ polarization. $\sigma$ = 2/3 is the cross phase modulation parameter. $\kappa_j$, $\kappa_{1j}$ and $\alpha_j$ are coupling coefficient, coupling coefficient dispersion and loss coefficient of the respective j$^{th}$ polarization respectively.

Followed with incorporating the optical parameters calculated using FEM in the CNLSEs, the individual pulse parameter dynamics are investigated by employing POM. The generalized POM has a phase constant $\theta$, which can take any value from -$\pi/2$ to $\pi/2$, that will determine, whether the Lagrangian is minimized corresponding to the nonlinear Schr\"{o}dinger equation (NLSE) or the residual field which is the difference between the assumed input profile and the exact pulse. Minimizing the Lagrangian of the NLSE is equivalent to the famous Lagrangian variational method \cite{Anderson} and minimizing the residual field is equal to the bare approximation of the CV theory \cite{Uthayakumar1,Nakkeeran1}. We assume $u_x$ and $u_y$ for the optical pulse through x and y polarized modes of core 1 and $v_x$ and $v_y$ through x- and y- polarized modes of the core 2, as a Gaussian function of the following form
\begin{eqnarray}
\label{Gaussian1}u_x=f_1(x_{1l}(z),t),\\
\label{Gaussian2}v_x=f_2(x_{2l}(z),t),\\
\label{Gaussian3}u_y=f_1(y_{1l}(z),t),\\
\label{Gaussian4}v_y=f_2(y_{2l}(z),t)\,\,.
\end{eqnarray}
where
\begin{eqnarray}
\label{pulseparametersx} f_{m}=x_{m1}\exp\left[-\frac{t^2}{x_{m2}^2}+\frac{i\,x_{m3}\,t^2}{2} +i\,x_{m4}\right],\\
\label{pulseparametersy} f_{n}=y_{n1}\exp\left[-\frac{t^2}{y_{n2}^2}+\frac{i\,y_{n3}\,t^2}{2}+i\,y_{n4}\right].\,\,
\end{eqnarray}
with $\ell=1,2,3,4$, $m=n=1,2$, the $x_{1\ell}$'s and $y_{1\ell}$'s designate the input beam parameters through the x- and y-polarized modes of the core 1 respectively, and $x_{2\ell}$'s and $y_{2\ell}$'s are the  input beam parameters through the x- and y-polarized modes of the core 2 respectively. The set of parameters ($x_{m1}$, $x_{m2}$, $x_{m3}$, $x_{m4}$) and ($y_{n1}$, $y_{n2}$, $y_{n3}$, and $y_{n4}$) represent the pulse amplitude, pulse width, chirp, and phase for x- and y- polarized modes respectively.

To proceed further, we substitute $Eqs.~(\ref{Gaussian1},\ref{Gaussian2},\ref{Gaussian3},\ref{Gaussian4})$ into $Eqs.~(\ref{CNLSE1},\ref{CNLSE2},\ref{CNLSE3},\ref{CNLSE4})$ and multiplying by the
projection  operators $P_{m}=f_{m\,x_{m \ell}}^{*} \,e^{i\theta}$ and $P_{n}=f_{n\,x_{n \ell}}^{*} \,e^{i\theta}$ for $Eqs.~(\ref{CNLSE1}\&\ref{CNLSE2})$  and $Eqs.~(\ref{CNLSE3}\&\ref{CNLSE4})$ respectively, where $\theta$ is a phase constant.  $f_{m\,x_{m l}}^{*}$ and $f_{n\,x_{nl}}^{*}$ represent complex conjugates of $f_{m\,x_{m \ell}}$ and $f_{n\,x_{n \ell}}$ respectively and $f_{m\,x_{m \ell}}$ and $f_{n\,x_{n \ell}}$ represents differentiation with respect to the argument $x_{m\ell}$ and $x_{n\ell}$ respectively. Integrating over t and collecting the real part, we get
\begin{eqnarray}
\label{Pom1}\nonumber-\int_{-\infty}^{\infty}\Im\left(f_{m\,z}\,f_{m\,x_{m \ell}}^{*}
\,e^{i\theta}\right)dt-\beta_{1x}\int_{-\infty}^{\infty}\Im\left(f_{m\,t}\,f_{m\,x_{m \ell}}^{*}
\,e^{i\theta}\right)dt\\-\frac{\beta_{2x}}{2}\int_{-\infty}^{\infty}\Re\left(f_{m\,tt}\, f_{m\,x_{m
\ell}}^{*}\,e^{i\theta}\right)dt+\kappa_x\int_{-\infty}^{\infty}\Re\left(f_{3-m}\,f_{m\,x_{m \ell}}^{*}\,e^{i\theta}\right)dt\nonumber\\+\int_{-\infty}^{\infty}\left(\gamma_x(|f_m|^2+{\sigma|f_{n}|^2})\right)\,\Re\left(f_m\,
f_{m\,x_{m \ell}}^{*} \,e^{i\theta}\right)dt\nonumber\\-\kappa_{1x}\int_{-\infty}^{\infty}\Im\left(f_{3-m\,t}\,f_{m\,x_{m \ell}}^{*}
\,e^{i\theta}\right)dt+\frac{\alpha_x}{2}\int_{-\infty}^{\infty}
\Im\left(f_m\,f_{m\,x_{m \ell}}^{*} \,e^{i\theta}\right)dt=0,\,\,\,\,\,\,\,\,
\\\label{Pom2} \nonumber -\int_{-\infty}^{\infty}\Im\left(f_{n\,z}\,f_{n\,x_{n \ell}}^{*}
\,e^{i\theta}\right)dt-\beta_{1y}\int_{-\infty}^{\infty}\Im\left(f_{n\,t}\,f_{n\,x_{n \ell}}^{*}
\,e^{i\theta}\right)dt\\\nonumber-\frac{\beta_{2y}}{2}\int_{-\infty}^{\infty}\Re\left(f_{n\,tt}\, f_{n\,x_{n
\ell}}^{*}\,e^{i\theta}\right)dt+\kappa_y\int_{-\infty}^{\infty}\Re\left(f_{3-n}\,f_{n\,x_{n \ell}}^{*}\,e^{i\theta}\right)dt\\\nonumber+\int_{-\infty}^{\infty}\left(\gamma_y(|f_n|^2+{\sigma|f_{m}|^2})\right)\,\Re\left(f_n\,
f_{n\,x_{n \ell}}^{*} \,e^{i\theta}\right)dt\nonumber\\ -\kappa_{1y}\int_{-\infty}^{\infty}\Im\left(f_{3-n\,t}\,f_{n\,x_{n \ell}}^{*}
\,e^{i\theta}\right)dt+\frac{\alpha_y}{2}\int_{-\infty}^{\infty}
\Im\left(f_n\, f_{n\,x_{n \ell}}^{*} \,e^{i\theta}\right)dt=0.\,\,\,\,\,\,\,\,
\end{eqnarray}
where $f_{mz},\,\,f_{nz}$ represents differentiation of $f_{m},\,\,f_{n}$ with respect to \textit{z}, and ($f_{mt},\,\,f_{nt}$), ($f_{mtt},\,\,f_{ntt}$) represents first and second order differentiation of $f_{m}$ and $f_{n}$ with respect to t respectively.

\noindent Proceeding with substituting Eqs.~(\ref{pulseparametersx}\,\&\,\ref{pulseparametersy}) into Eqs.~(\ref{Pom1}\,\&\,\ref{Pom2}) and choosing any value for $\theta$
between $-\pi/2$ to $\pi/2$, we obtain the collective variable (CV) equations of motion for $u_x(z,t)$, $v_x(z,t)$, $u_y(z,t)$ and $v_y(z,t)$ as follows
\begin{eqnarray}
\label{odex1}
\dot{x}_{m1}=\frac{1}{4\sqrt{2\pi}}\bigg(2\,x_{m1}\bigg(\frac{2\,\beta_{1x}}{x_{m2}}+\sqrt{2\pi}(\alpha_x
+x_{m3}\,\beta_{2x})\bigg)+\frac{1}{\rho_{11}}\,4\,x_{m2}\,x_{n1}\,x_{n2}^2(4\,\kappa_{1x}\,\nonumber\\(-8\,\cos(\rho_5)
+3\,\sqrt{\rho_9}\,x_{m2}^2\,\cos(\rho_7))+x_{n2}^2\,(\sqrt{2\,\pi}\,\rho_9^{1/4}\,\kappa_x\,(\delta_{\mp}\,4\,sin(\rho_1)
+\delta_{\pm}\,3\nonumber\\\,\sqrt{\rho_9}\,x_{m2}^2\,sin(\rho_3))+2\,x_{m3}\,\kappa_{1x}\,(\delta_{\pm}\,
8\,sin(\rho_5)-\delta_{\mp}\,3\,\sqrt{\rho_9}\,x_{m2}^2)sin(\rho_7)))\bigg),\,\,\,\,\,\,\,\,
\end{eqnarray}
\begin{eqnarray}
\label{odex2}
\nonumber\dot{x}_{m2}=\frac{1}{x_{m1}\,\rho_{11}}\bigg(x_{m1}\bigg(\sqrt{\frac{2}{\pi}}\,\beta_{1x}
-x_{m2}\,x_{m3}\,\beta_{2x}\bigg)\rho_{11}+2\,x_{m2}^2\,x_{n1}\,x_{n2}^2\bigg(16\,\sqrt{\frac{2}{\pi}}
\,\kappa_{1x}\\cos(\rho_5)
\nonumber-2\,\sqrt{\frac{2}{\pi}}\,x_{m2}^2\,\kappa_{1x}\,\sqrt{\rho_9}\,\cos(\rho_7)+x_{n2}^2\bigg(\delta_{\pm}
\,4\,\kappa_x\,\rho_9^{1/4}\,sin(\rho_1)+\delta_{\mp}\,x_{m2}^2\,\\\kappa_x\,\rho_9^{3/4}\,sin(\rho_3)
+\sqrt{\frac{2}{\pi}}\,x_{n3}\,\kappa_{1x}\,(\delta_{\mp}\,8\,sin(\rho_5)+\delta_{\pm}\,x_{m2}^2\,\sqrt{\rho_9}
\,sin(\rho_7))\bigg)\bigg)\bigg),\,\,\,\,\,\,\,\,\,
\end{eqnarray}
\begin{eqnarray}
\label{odex3}
\nonumber\dot{x}_{m3}=-\frac{1}{\rho_{11}\,\eta_{ma}^2\,x_{m1}\,x_{m2}^4}\bigg(\eta_{ma}^2\,\bigg(\rho_{11}
\,x_{m1}\,\bigg((4-x_{m2}^4\,x_{m3}^2)\,\beta_{2x}+\sqrt{2}\,x_{m2}^2\,\\\nonumber\bigg(\frac{2\,x_{m2}\,x_{m3}
\,\beta_{1x}}{\sqrt{\pi}}+x_{m1}^2\,\gamma_x\bigg)\bigg)+8\,x_{m2}^3\,x_{n1}\,x_{n2}^2\,\bigg(-4\rho_9^{1/4}
\,x_{n2}^2\,\kappa_x\,\cos(\rho_1)\\\nonumber+\rho_9^{3/4}\,x_{m2}^2\,x_{n2}^2\,\kappa_x\,\cos(\rho_3)
+\sqrt{\frac{2}{\pi}}\,\kappa_{1x}(8\,x_{n2}^2\,x_{n3}\,\cos(\rho_5)+\delta_{\pm}16\,sin(\rho_5)\\\nonumber
+\delta_{\mp}\sqrt{\rho_9}\,x_{m2}^2\,(\delta_{\pm}x_{n2}^2\,x_{n3}\,\cos(\rho_7)+2sin(\rho_7)))\bigg)\bigg)
+4\,\rho_{11}\,\sqrt{\eta_{mb}}\,x_{m1}\\\,x_{m2}^5\,\gamma_x\,\sigma\,y_{m1}^2\,y_{m2}^2\bigg),
\end{eqnarray}
\begin{eqnarray}
\label{odex4}
\nonumber\dot{x}_{m4}=\frac{1}{8\,\sqrt{\pi}\,\rho_{11}\,\eta_{ma}^2\,x_{m1}\,x_{m2}^2}\bigg(-2\sqrt{2}
\,\,\eta_{ma}^2\,x_{m2}^3\bigg(\rho_{11}\,x_{m1}\,x_{m3}\,\beta_{1x}+4\,x_{n1}\\\nonumber\,x_{n2}^2\,\kappa_{1x}
\,\bigg(-8\,x_{n2}^2\,x_{n3}\,\cos(\rho_5)+\delta_{\mp}\,16\,sin(\rho_5)+3\,\sqrt{\rho_9}
\,x_{m2}^2\,(x_{n2}^2\,x_{n3}\,\\\nonumber \cos(\rho_7)+\delta_{\pm}\,2\,sin(\rho_7))\bigg)\bigg)
+\sqrt{\pi}\,\bigg(\eta_{ma}^2\,\bigg(\rho_{11}\,x_{m1}(8\,\beta_{2x}+5\,\sqrt{2}\,x_{m1}^2\\\nonumber\,x_{m2}^2
\,\gamma_x)+8\,\rho_9^{1/4}\,x_{m2}^3\,x_{n1}\,x_{n2}^4\,\kappa_x\,(-4\,\cos(\rho_1)+3\,\sqrt{9}\,x_{m2}^2
\,\cos(\rho_3))\bigg)\\+4\,\rho_{11}\,\sqrt{\eta_{mb}}\,x_{m1}\,x_{m2}^3\,\gamma_x\,\sigma\,y_{m1}^2\,y_{m2}^2
(3x_{m2}^2+2y_{m2}^2)\bigg)\bigg).\,\,
\end{eqnarray}
\begin{eqnarray}
\label{odey1}
\nonumber\dot{y}_{n1}=\frac{1}{4\sqrt{2\pi}}\bigg(2\,y_{n1}\bigg(\frac{2\,\beta_{1y}}{y_{n2}}+\sqrt{2\pi}(\alpha_y
+y_{n3}\,\beta_{2y})\bigg)+\frac{1}{\rho_{12}}\,4\,y_{n2}\,y_{m1}\,y_{m2}^2(4\,\kappa_{1y}\,
\\\nonumber(-8\,\cos(\rho_6)+3\,\sqrt{\rho_{10}}\,y_{n2}^2\,\cos(\rho_8))+y_{m2}^2\,(\sqrt{2\,\pi}\,\rho_{10}^{1/4}\,
\kappa_y\,(\delta_{\mp}\,4\,sin(\rho_2)+\delta_{\pm}\,3\\\,\sqrt{\rho_{10}}\,y_{n2}^2\,sin(\rho_4))
+2\,y_{m3}\,\kappa_{1y}\,(\delta_{\pm}\,8\,sin(\rho_6)-\delta_{\mp}\,3\,\sqrt{\rho_{10}}\,y_{n2}^2)
\,sin(\rho_8)))\bigg),\,\,\,\,\,\,\,\,
\end{eqnarray}

\begin{eqnarray}
\label{odey2}
\nonumber\dot{y}_{n2}=\frac{1}{y_{n1}\,\rho_{12}}\bigg(y_{n1}\bigg(\sqrt{\frac{2}{\pi}}\,\beta_{1y}
-y_{n2}\,y_{n3}\,\beta_{2y}\bigg)\rho_{12}+2\,y_{n2}^2\,y_{m1}\,y_{m2}^2\bigg(16\,\sqrt{\frac{2}{\pi}}
\,\kappa_{1y}\\\nonumber\,\cos(\rho_6)-2\,\sqrt{\frac{2}{\pi}}\,y_{n2}^2\,\kappa_{1y}\,\sqrt{\rho_{10}}
\,\cos(\rho_8)+y_{m2}^2\bigg(\delta_{\pm}\,4\,\kappa_y\,\rho_{10}^{1/4}\,sin(\rho_2)+\delta_{\mp}\,y_{n2}^2
\,\\\kappa_y\,\rho_{10}^{3/4}\,sin(\rho_4)+\sqrt{\frac{2}{\pi}}\,y_{m3}\,\kappa_{1y}\,(\delta_{\mp}\,8\,sin(\rho_6)
+\delta_{\pm}\,y_{n2}^2\,\sqrt{\rho_{10}}\,sin(\rho_8))\bigg)\bigg)\bigg),\,\,\,\,
\end{eqnarray}

\begin{eqnarray}
\label{odey3}
\nonumber\dot{y}_{n3}=-\frac{1}{\rho_{12}\,\eta_{na}^2\,y_{n1}\,y_{n2}^4}\bigg(\eta_{na}^2\,\bigg(\rho_{12}
\,y_{n1}\,\bigg((4-y_{n2}^4\,y_{n3}^2)\,\beta_{2y}+\sqrt{2}\,y_{n2}^2\,\,\\\nonumber\,\,\bigg(\frac{2\,y_{n2}\,y_{n3}
\,\beta_{1y}}{\sqrt{\pi}}+y_{n1}^2\,\gamma_y\bigg)\bigg)+8\,y_{n2}^3\,y_{m1}\,y_{m2}^2\,
\bigg(-4\rho_{10}^{1/4}\,y_{m2}^2\,\kappa_y\,\cos(\rho_2)\\\nonumber+\rho_{10}^{3/4}\,y_{n2}^2\,y_{m2}^2\,\kappa_y
\,\cos(\rho_4)+\sqrt{\frac{2}{\pi}}\,\kappa_{1y}(8\,y_{m2}^2\,y_{m3}\,\cos(\rho_6)\\\nonumber+\delta_{\pm}16
\,sin(\rho_6)+\delta_{\mp}\sqrt{\rho_{10}}\,y_{n2}^2\,(\delta_{\pm}\,y_{m2}^2\,y_{m3}\,\cos(\rho_8)
+2sin(\rho_8)))\bigg)\bigg)\\+4\,\rho_{12}\,\sqrt{\eta_{nb}}\,y_{n1}\,y_{n2}^5\,\gamma_y\,\sigma\,x_{n1}^2
\,x_{n2}^2\bigg),
\end{eqnarray}

\begin{eqnarray}
\label{odey4}
\nonumber\dot{y}_{n4}=\frac{1}{8\,\sqrt{\pi}\,\rho_{12}\,\eta_{na}^2\,y_{n1}\,y_{n2}^2}\bigg(-2\sqrt{2}\,\,\eta_{na}^2
\,y_{n2}^3\bigg(\rho_{12}\,y_{n1}\,y_{n3}\,\beta_{1y}+4\,y_{m1}\,y_{m2}^2\,\kappa_{1y}\\\nonumber\,\bigg(-8\,y_{m2}^2
\,y_{m3}\,\cos(\rho_6)+\delta_{\mp}\,16\,sin(\rho_6)+3\,\sqrt{\rho_{10}}\,y_{n2}^2\,(y_{m2}^2\,y_{m3}\,\cos(\rho_8)
\\\nonumber+\delta_{\pm}\,2\,sin(\rho_8))\bigg)\bigg)+\sqrt{\pi}\,\bigg(\eta_{na}^2\,\bigg(\rho_{12}
\,y_{n1}(8\,\beta_{2y}+5\,\sqrt{2}\,y_{n1}^2\,y_{n2}^2\,\gamma_y)\\\nonumber+8\,\rho_{10}^{1/4}\,y_{n2}^3\,y_{m1}\,y_{m2}^4
\,\kappa_y\,(-4\,\cos(\rho_2)+3\,\sqrt{9}\,y_{n2}^2\,\cos(\rho_4))\bigg)\\+4\,\rho_{12}\,\sqrt{\eta_{nb}}\,y_{n1}
\,y_{n2}^3\,\gamma_y\,\sigma \,x_{n1}^2\,x_{n2}^2(3y_{n2}^2+2x_{n2}^2)\bigg)\bigg).\,\,\,\,\,\,\,\,\,\,
\end{eqnarray}
\noindent where\\
$\rho_1=\,\frac{3}{2}\cot^{-1}\bigg[\frac{2(x_{12}^2+x_{22}^2)}{x_{12}^2\,x_{22}^2(x_{13}-x_{23})}\bigg]+x_{14}-x_{24}$,
$\rho_2=\,\frac{3}{2}\cot^{-1}\bigg[\frac{2(y_{12}^2+y_{22}^2)}{y_{12}^2\,y_{22}^2(y_{13}-y_{23})}\bigg]+y_{14}-y_{24}$,\\
$\rho_3=\,\frac{1}{2}\cot^{-1}\bigg[\frac{2(x_{12}^2+x_{22}^2)}{x_{12}^2\,x_{22}^2(x_{13}-x_{23})}\bigg]+x_{14}-x_{24}$,
$\rho_4=\,\frac{1}{2}\cot^{-1}\bigg[\frac{2(y_{12}^2+y_{22}^2)}{y_{12}^2\,y_{22}^2(y_{13}-y_{23})}\bigg]+y_{14}-y_{24}$,\\
$\rho_5=\,2\cot^{-1}\bigg[\frac{2(x_{12}^2+x_{22}^2)}{x_{12}^2\,x_{22}^2(x_{13}-x_{23})}\bigg]+x_{14}-x_{24}$,
$\rho_6=\,2\cot^{-1}\bigg[\frac{2(y_{12}^2+y_{22}^2)}{y_{12}^2\,y_{22}^2(y_{13}-y_{23})}\bigg]+y_{14}-y_{24}$,\\
$\rho_7=\,\cot^{-1}\bigg[\frac{2(x_{12}^2+x_{22}^2)}{x_{12}^2\,x_{22}^2(x_{13}-x_{23})}\bigg]+x_{14}-x_{24}$,
$\rho_8=\,\cot^{-1}\bigg[\frac{2(y_{12}^2+y_{22}^2)}{y_{12}^2\,y_{22}^2(y_{13}-y_{23})}\bigg]+y_{14}-y_{24}$,\\
$\rho_9=\,\frac{4}{x_{12}^4}+\frac{4}{x_{22}^4}+\frac{8}{x_{12}^2x_{22}^2}+(x_{13}-x_{23})^2$,
$\rho_{10}=\,\frac{4}{y_{12}^4}+\frac{4}{y_{22}^4}+\frac{8}{y_{12}^2y_{22}^2}+(y_{13}-y_{23})^2$,\\
$\rho_{11}=\,8\,x_{12}^2\,x_{22}^2+4x_{22}^4+x_{12}^4(4+x_{22}^4(x_{13}-x_{23})^2)$,\\
$\rho_{12}=\,8\,y_{12}^2\,y_{22}^2+4y_{22}^4+y_{12}^4(4+y_{22}^4(y_{13}-y_{23})^2)$,\\
$\rho_{13}=\frac{1}{x_{12}^2}+\frac{1}{y_{12}^2}$,\,
$\rho_{14}=x_{12}^2+y_{12}^2$,\,
$\rho_{15}=\frac{1}{x_{22}^2}+\frac{1}{y_{22}^2}$,\,
$\rho_{16}=x_{22}^2+y_{22}^2$,\\
$\eta_{ma}=\rho_{14}^2,\,\rho_{16}^2$\,for $m=1,2$;\,
$\eta_{mb}=\rho_{13}^2,\,\rho_{15}^2$\,for $m=1,2$;\\
$\eta_{na}=\rho_{14}^2,\,\rho_{16}^2$\,for $n=1,2$;\,\,\,
$\eta_{nb}=\rho_{13}^2,\,\rho_{15}^2$\,for $n=1,2$,\\
$\delta_{\pm}=+1$ for $m=n=1$;\,\,\,\,\,\,\,\,\,\,$\delta_{\pm}=-1$ for $m=n=2$;\\$\delta_{\mp}=-1$ for $m=n=1$;\,\,\,\,\,\,\,\,\,\,$\delta_{\mp}=+1$ for $m=n=2$;\\
Here m and n takes the value 1, 2 (for core 1 and core2 for x- and y- polarized modes respectively) and $m \neq n$. The overdot denotes the differentiation with respect to \textit{z}.
\section{Coupling characteristics of CPCFC}
Coupling characteristics are investigated by varying the polarization angle of an input Gaussian pulse with an amplitude of 0.0024 W, pulse width of 10 ps at 1.55 $\mu$m for a length of 6x10$^{-3}$ m. Initial pulse of Gaussian form for the x and y polarized modes of the CPCFC is introduced through the core 1 and 0.0001 $\%$ of the input power through core 1 is assumed as input pulse for core 2. The Gaussian pulse through the cores in presence of polarization angle  is assumed as follows \cite{Li2}
\begin{eqnarray}
\label{initialpulse}
u_x(0,t)=A_0\,\cos(\theta)\,\exp(-t^2/W_0),\\
u_y(0,t)=A_0\,\sin(\theta)\,\exp(-t^2/W_0),\\
v_x(0,t)=0.0001\% \,\texttt{of}\, u_x(0,t),\\
v_y(0,t)=0.0001\% \,\texttt{of}\, u_y(0,t).
\end{eqnarray}
where $A_0$ is the input amplitude, $W_0$ is the pulse width and $\theta$ is the polarization angle provided with the values: ($\pi$/12, $\pi$/4, $\pi$/3 $\&$ $\pi$/2). To explore the coupling characteristics, first we consider the case $\theta$ = $\pi$/4, for which input pulse exhibits both the polarization components with equal strength. And next case, the influence of polarization angle on dynamics of the system is studied by means of varying $\theta$.
\begin{figure}[tbp]
\label{amplitudex}
\begin{center}
\includegraphics[width=0.485\textwidth]{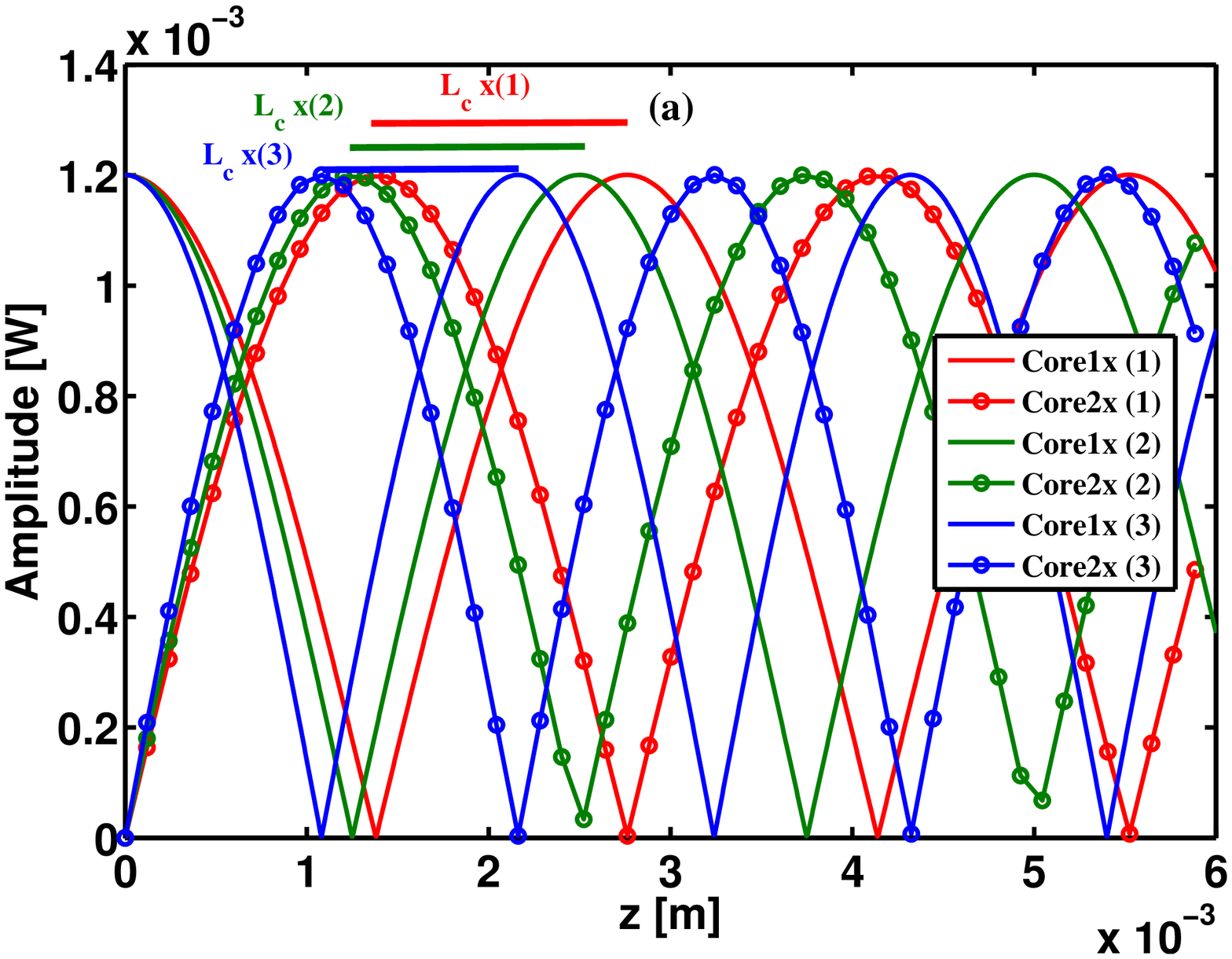}
\label{amplitudey}
\includegraphics[width=0.485\textwidth]{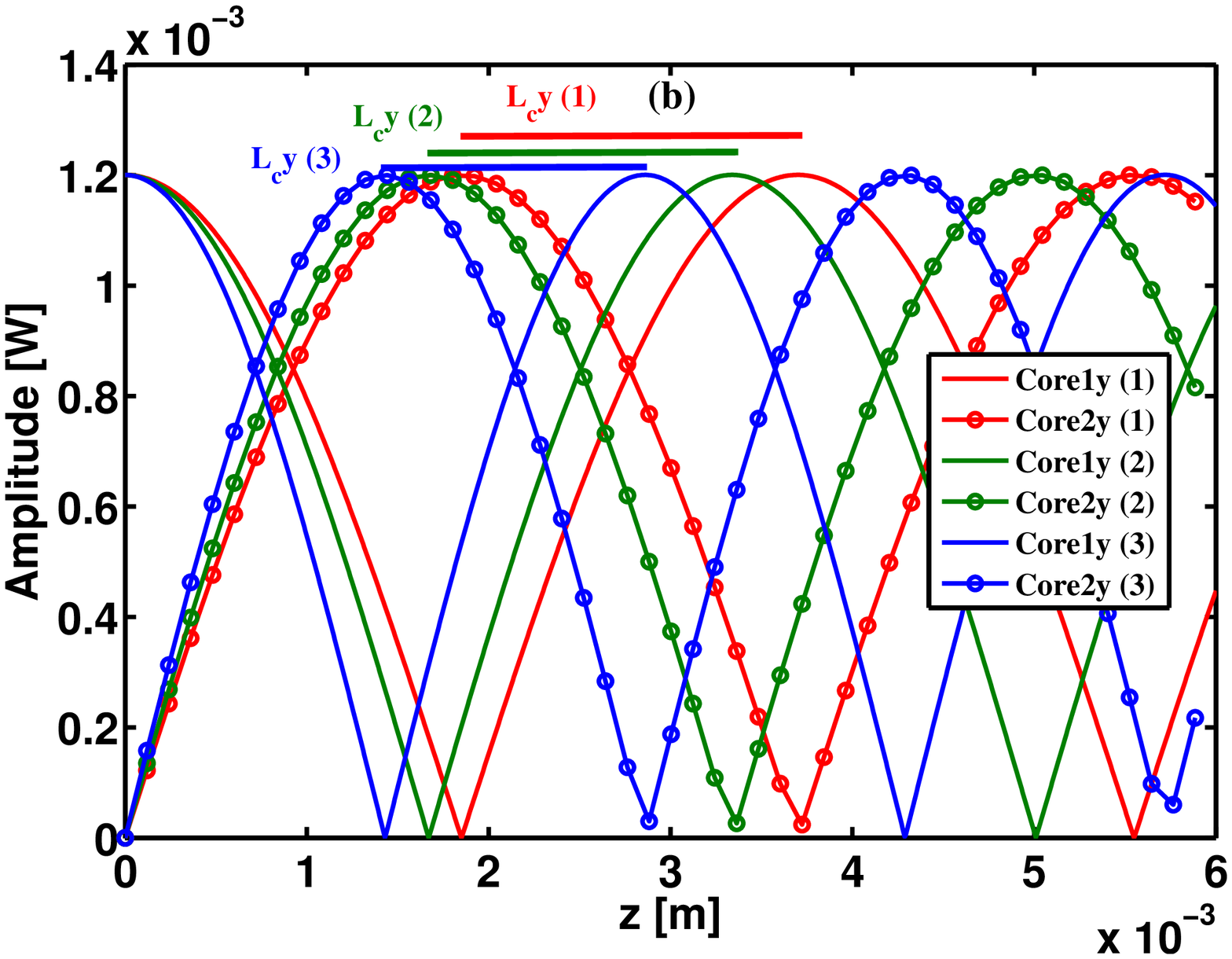}
\end{center}
\caption{Evolution of the amplitude as a function of the distance for (a) x- and (b) y- polarization.}
\end{figure}
\begin{figure}[tbp]
\label{pulsewidthx}
\begin{center}
\includegraphics[width=0.485\textwidth]{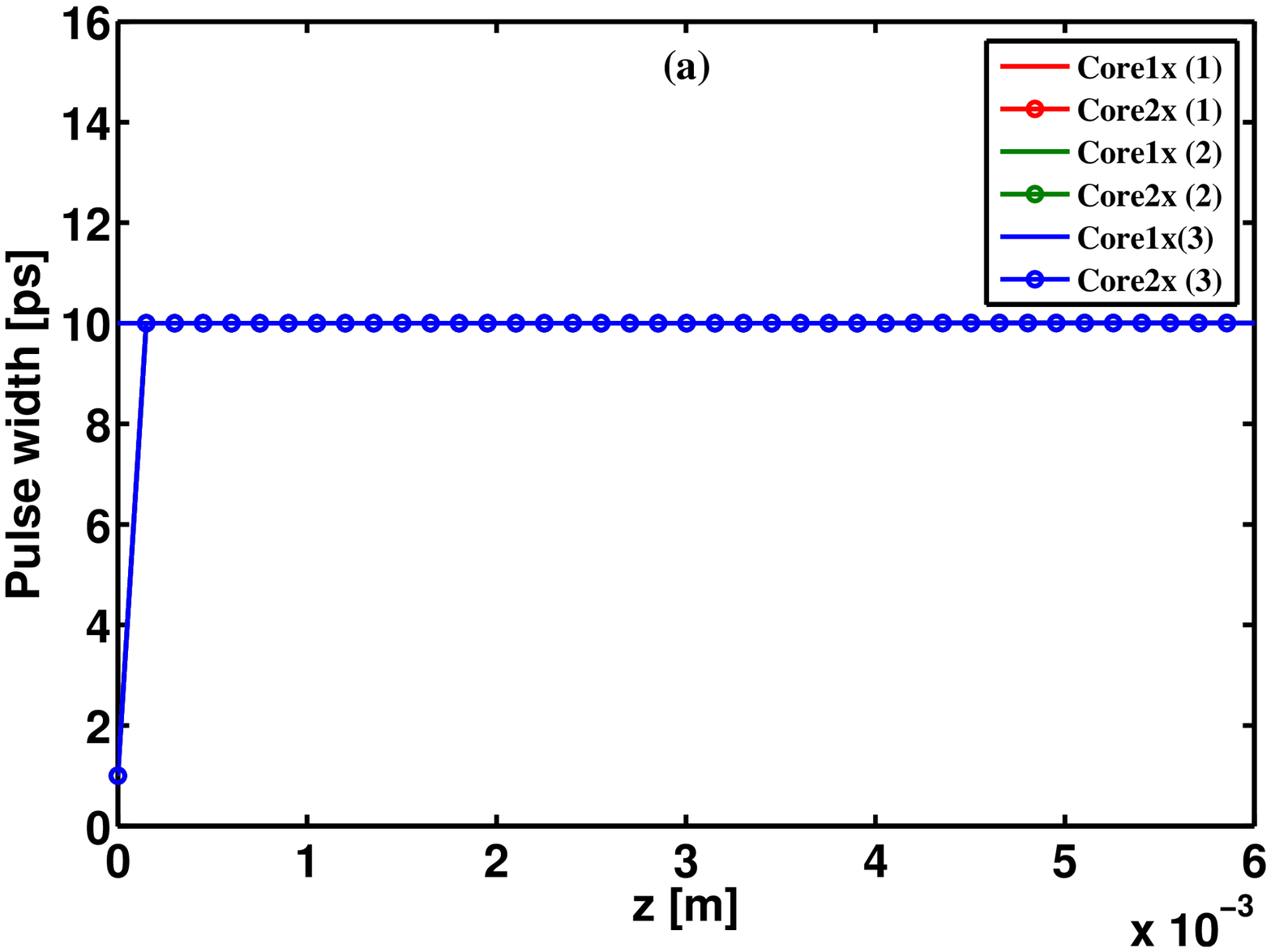}
\label{pulsewidthy}
\includegraphics[width=0.485\textwidth]{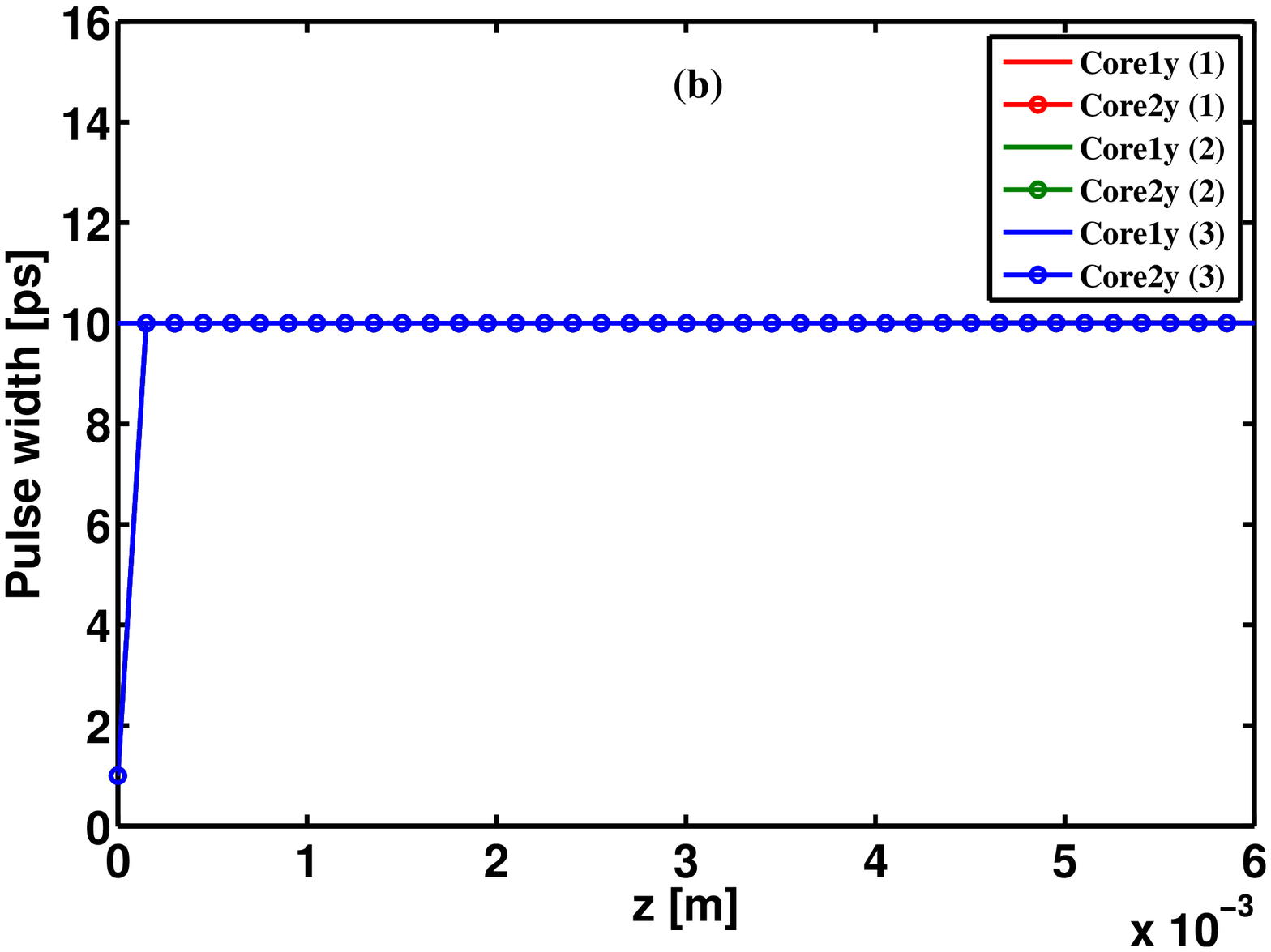}
\end{center}
\caption{Evolution of the pulse width as a function of the distance for (a) x- and (b) y- polarization.}
\end{figure}
\subsection{Case (i): $\theta\,=\,\pi/4$}
The evolution of amplitude as the function of the distance for x- and y- polarized components with polarization angle $\pi$/4 for input Gaussian pulse are illustrated by Fig. 5 (a) and Fig. 5 (b) respectively. Red, green and blue lines show the pulse propagation through core 1 and the red, green and blue lines with dot shows the pulse propagation through core 2 for the designs 1, 2 and 3 respectively for both the components. When an input pulse is introduced through the core 1 at z = 0, the amplitude is maximum through core 1 indicated by the lines without dot and it is minimum for core 2 shown by dotted lines. As the pulse propagates and reaches the distance equals to the L$_c$ the input power is completely transferred to the neighboring core at respective L$_c$ with a phase shift of $\pi/2$. As the L$_c$ of the design 1 is greater than the other two designs, pulse is subjected to travel a longer distance to reach the core 2. And number of oscillations of pulse between the cores increases as one move from design 1 to 3 due to reduction in L$_c$. The splitting of the x- and y- polarized components of the individual designs increase with the arrival of subsequent L$_c$s. Moreover, the distance between the polarized components is greater for the design with greater L$_c$. For both the components, the amplitude remains constant for all the designs throughout the propagation. Thus, the coupler exhibits tuned coupling even in presence of asymmetry in the core for birefringent axes at linear regime and remains robust to the asymmetry throughout propagation.
\begin{figure}[tbp]
\label{chirpx}
\begin{center}
\includegraphics[width=0.48\textwidth]{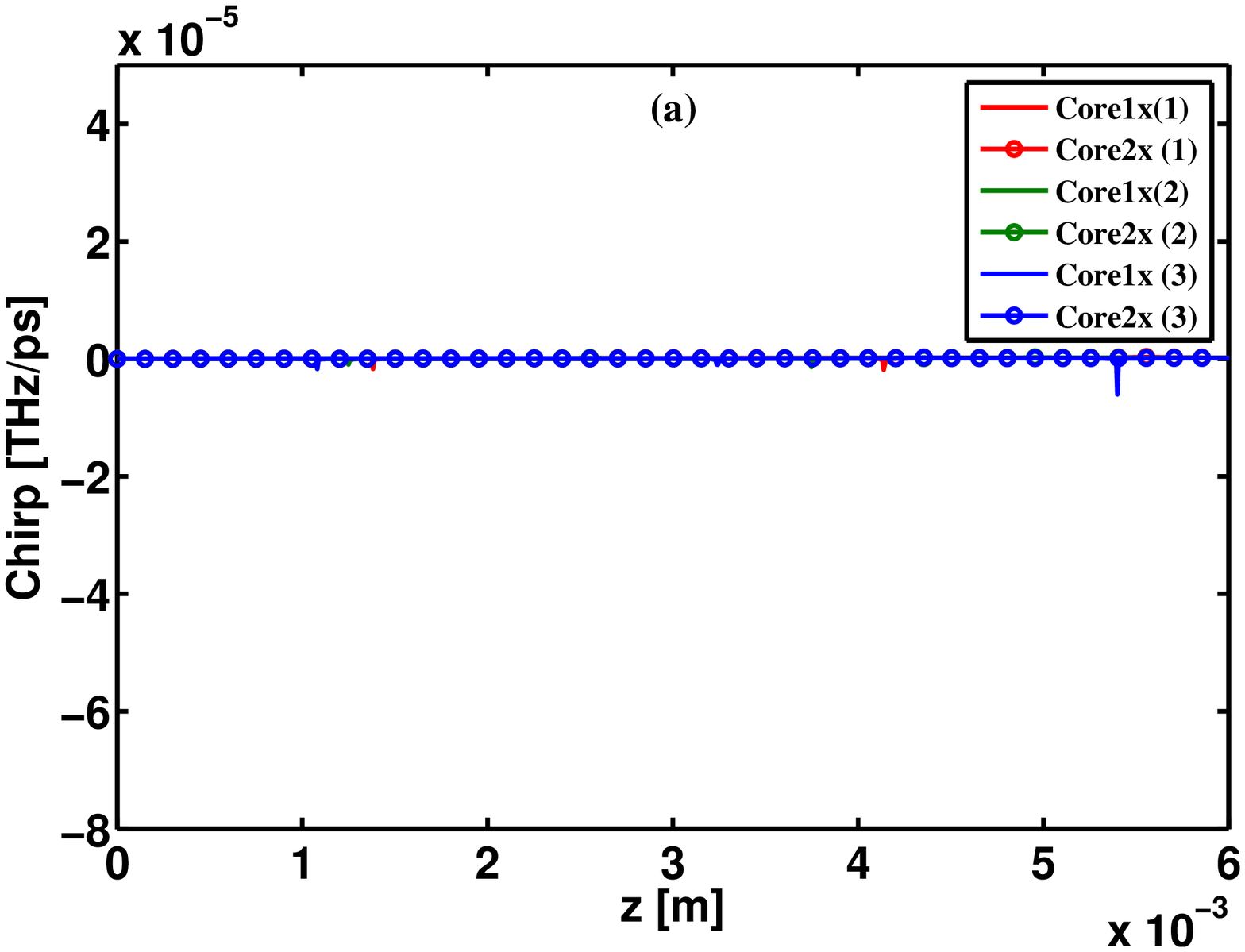}
\label{chirpy}
\includegraphics[width=0.48\textwidth]{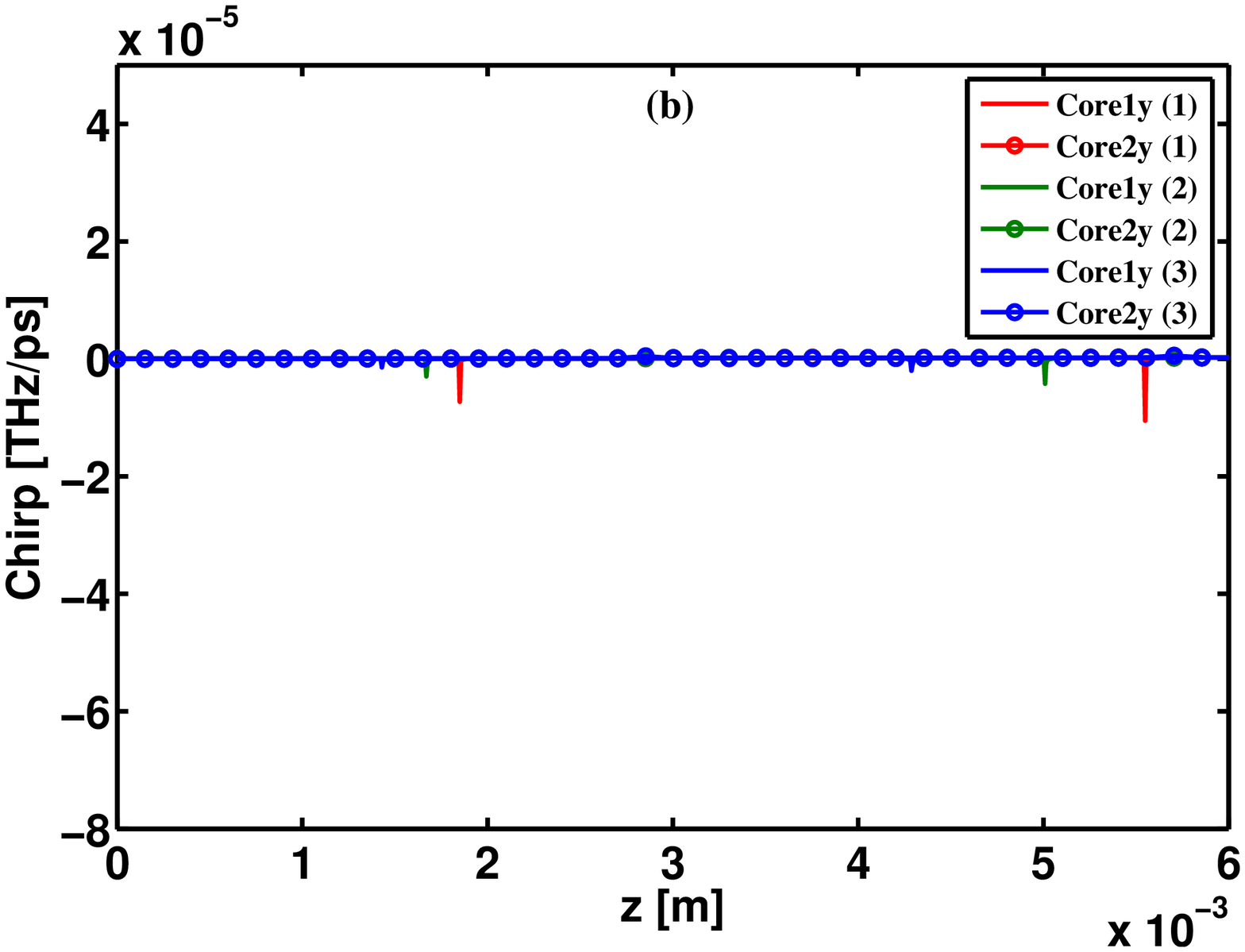}
\end{center}
\caption{Evolution of the chirp as a function of the distance for (a) x- and (b) y- polarization.}
\end{figure}
\begin{figure}[tbp]
\label{phasex}
\begin{center}
\includegraphics[width=0.48\textwidth]{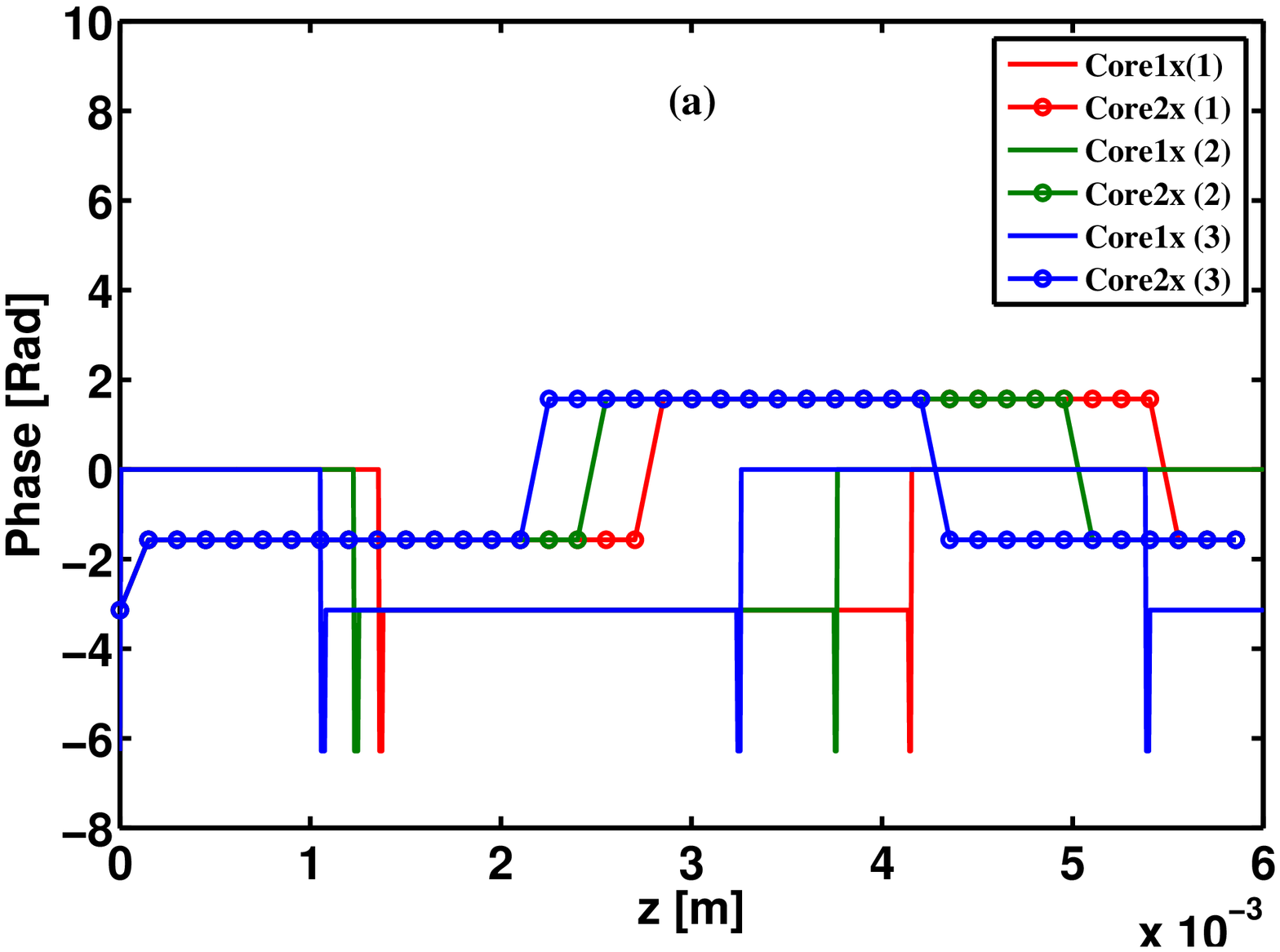}
\label{phasey}
\includegraphics[width=0.48\textwidth]{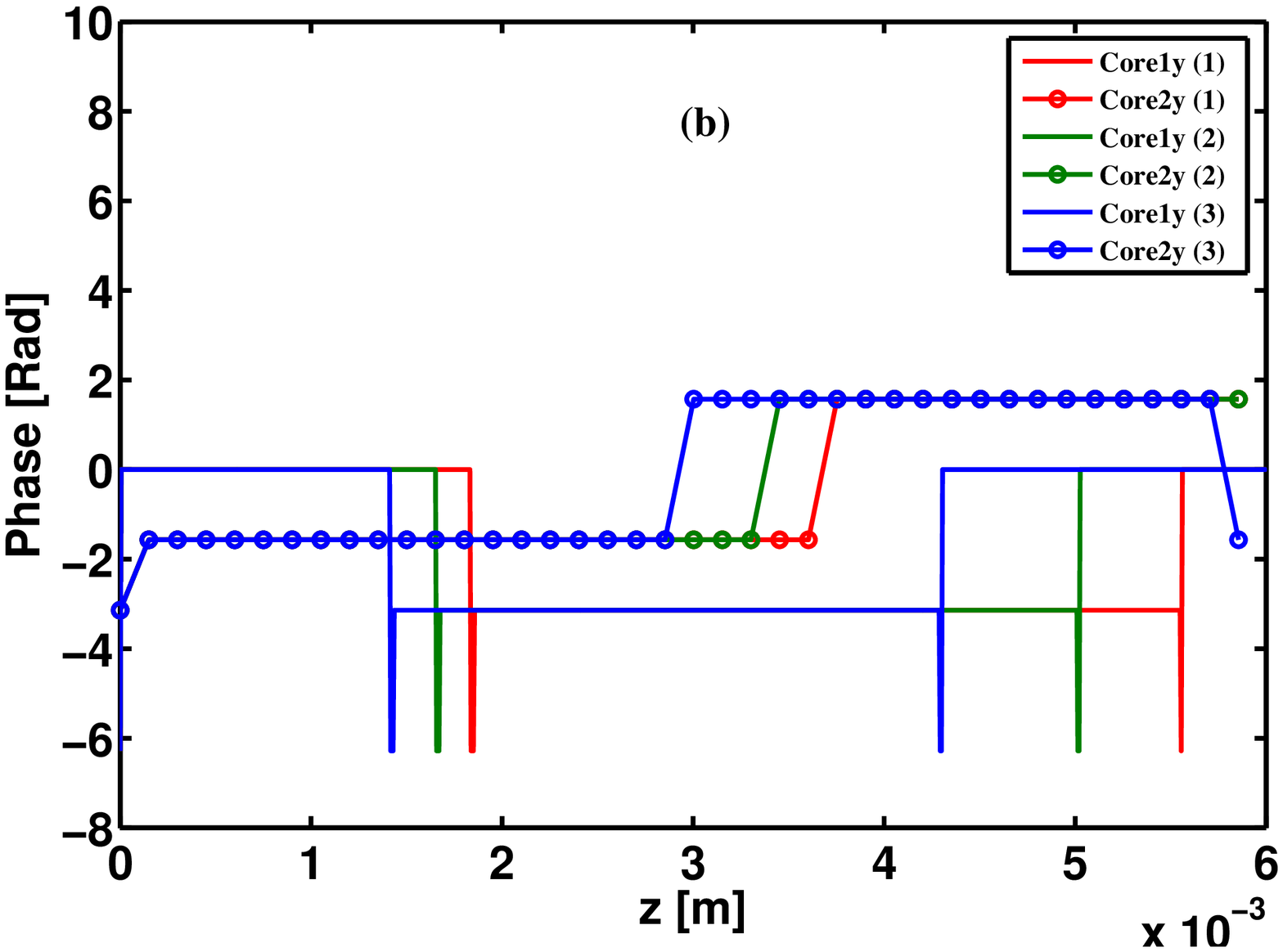}
\end{center}
\caption{Evolution of the phase as a function of the distance for (a) x- and (b) y- polarization.}
\end{figure}

Figs. 6 (a $\&$ b) depicts the variation of pulse width as a function of the distance for x- and y- polarized components respectively. Initial pulse width provided for core 1 is 10 ps and that of core 2 is assumed as 1 ps. As the pulse propagates the pulse width of the both cores becomes the same and remains constant throughout the propagation for both the components and for all the structures. This undistorted pulse propagation throughout the propagation length is due to the greater dispersion length (L$_D$ $\simeq$ 1.1 km) as well as nonlinear length (L$_{NL}$ $\simeq$ 0.9 km). As a result, the system becomes robust to the pulse break up due to shorter L$_c$s compared to that of the   L$_D$ as well as L$_{NL}$. The variation of chirp as the function of the distance for birefringent components are illustrated by Figs. 7 (a $\&$ b). Frequency chirping is absent during the propagation and it remains constant throughout of the propagation distance due to the higher value of walk-off length. Thus, all the frequency components remains intact during the propagation for both the components. As the pulse travels from modes of one core to other, after the arrival of every coupling it experiences a small perturbation. This perturbation appears as small spikes at the points of energy transfer. Thus, the spikes appear in the frequency chirp plots due to distortion of energy in the numerical simulation at the points of energy transfer from one core to the other, hence it is neglected in the analysis. Moreover, there is no significant contribution of CCD to pulse parameter dynamics due to its very low value. Hence, the influence of CCD in system dynamics is absent throughout the propagation.

The periodicity of the coupler as a function of the distance for birefringent components are portrayed by
Figs. 8 (a $\&$ b). Whenever an optical pulse is introduced through any of the input port of the coupler, it couples
to the neighboring port for the every arrival of the L$_c$ with a phase shift of $\pi/2$. This periodicity can be
viewed through the variation of phase as a function of distance. Moreover, the lag of phase with decreasing L$_c$
from design 1 to 3 can also be clearly viewed. In phase plot also, spikes appear due to energy distortion at the
regions of energy transfer and we neglected it in our analysis.
\begin{figure}[tbp]
\label{xpoltheta1}
\includegraphics[width=0.53\textwidth]{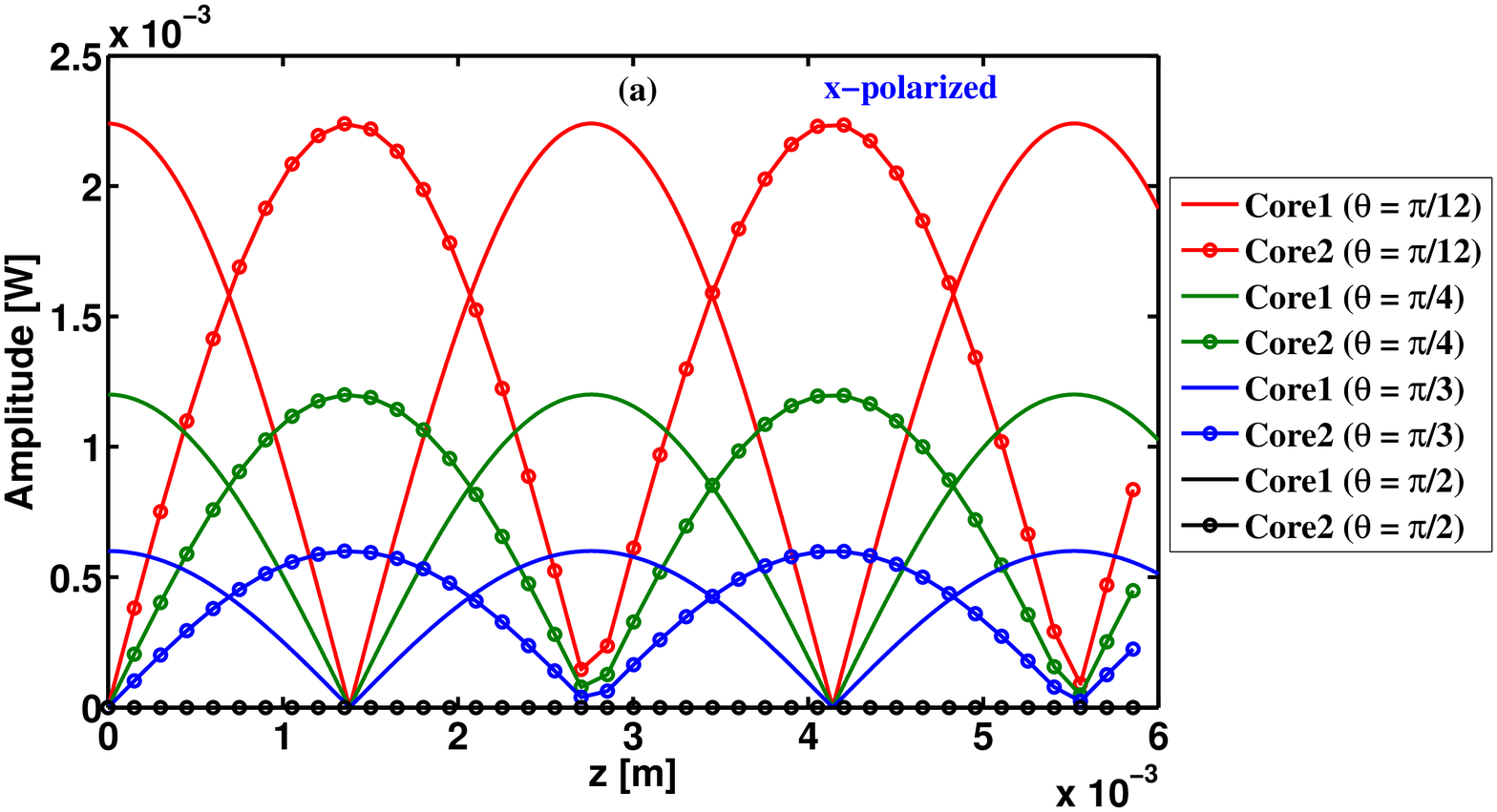}
\label{phasey}
\includegraphics[width=0.53\textwidth]{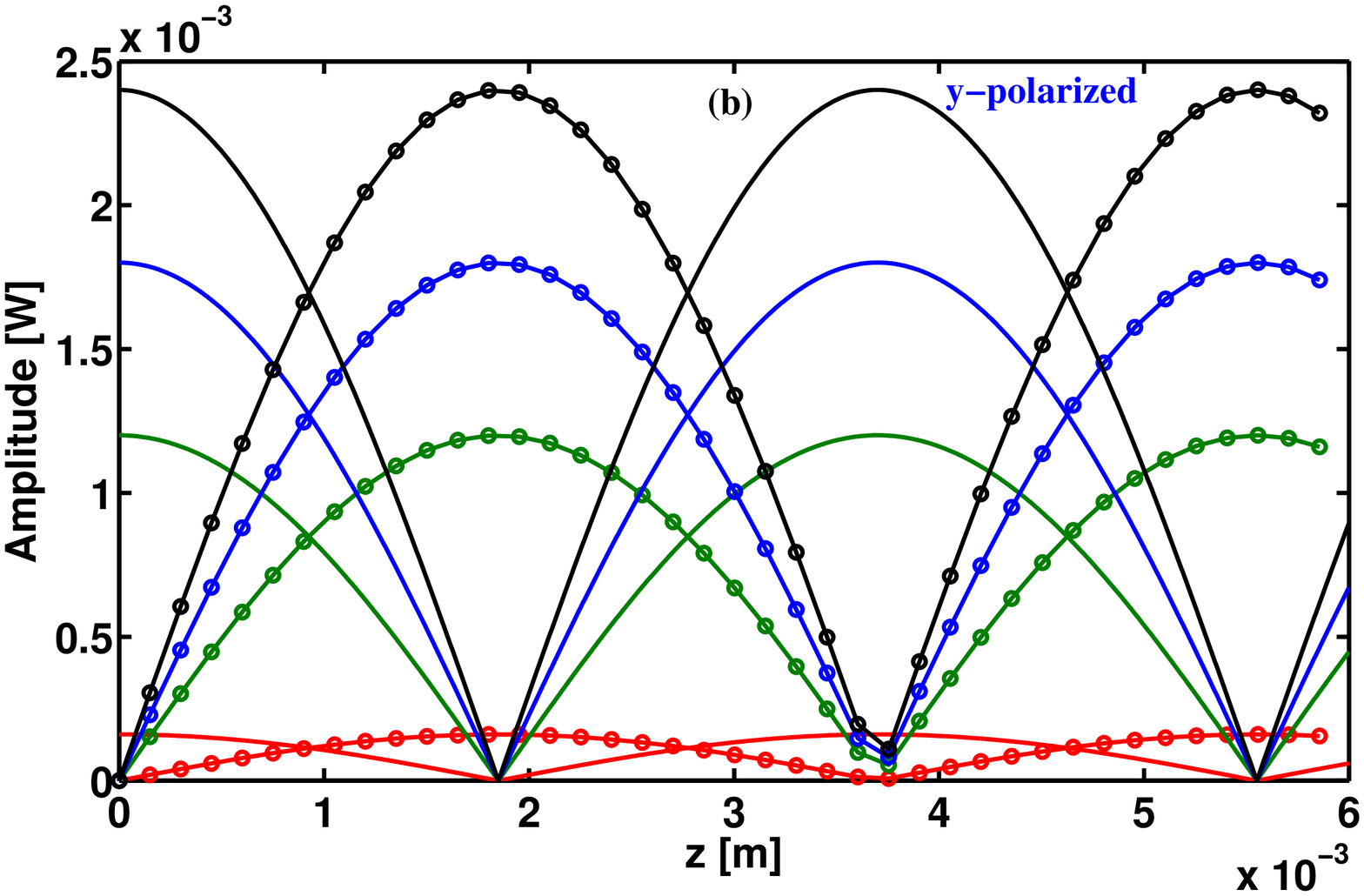}
\caption{Evolution of the amplitude as a function of the distance for different polarizing angle for (a) x- and (b) y- polarization.}
\end{figure}
\subsection{Case (ii): Varying $\theta$}
The dynamics of amplitude as the function of distance for different polarizing angle is illustrated by Figs. 9 (a $\&$ b) for x- and y- polarization for design 1. The polarizing angles chosen for our studies are  $\pi/12$, $\pi/4$, $\pi/3$ and $\pi/2$ indicated by the red, green, blue and black lines respectively. The line without dot display the pulse propagation through the core 1 and line with dot portrays the pulse propagation through core 2. When the input pulse is introduced with the polarizing angle $\theta$ = $\pi/12$, the polarization component exist is x-polarized and that of y-polarized becomes negligible. Thus, the amplitude of the x-polarized component becomes maximum and that of the y-polarized component becomes minimum as indicated by red lines of Figs. 9 (a $\&$ b). As the $\theta$ value of the input pulse is increased to $\pi/4$, then both the polarization component of the pulse exists with equal values. Thus, both the cores and both the birefringent components exhibits the same amplitude throughout the propagation as shown by the green lines of the Figs. 9 (a $\&$ b). For further increase in $\theta$ = $\pi/3$, the principal birefringent component exist is y polarization and that of x polarization component will becomes lesser. As a result, the amplitude through y-polarized mode increases and that through x-polarized mode decreases. Moreover, when the pulse is introduced with $\theta$ = $\pi/2$, the only polarization component that exists is y-polarization. Thus, the amplitude of the y-polarized component becomes maximum and that of x-polarized component becomes minimum shown by the maximum amplitude of the black line in Fig. 9 (b) and minimum amplitude of the black line in Fig. 9 (a). For other designs 2 $\&$ 3 also, the system demonstrates the same coupling characteristics with a little variation in amplitude. Hence, we restricted our discussions with a single design and which is sufficient to explain the characteristics of all the designs. In addition, to the variation in amplitude, there also exists an asymmetry in the value of amplitude for varying polarization angle at low input power level. This variation in amplitude along the birefringent axes is primarily due to the difference in optical properties along x- and y- polarized axes of the CPCFC. Thus, the polarization angle also influences the energy transfer through the birefringent modes. Thus, the input pulse with selective polarizing angle will allow us to control the splitting ratio as well as determine the existence of desired polarization component. These characteristics of the coupler make them suitable for the polarization splitting and sensing applications.
\section{Conclusion}
In this article, we have investigated the influence of diameter of the air hole and the role of polarizing angle of the input pulse on coupling dynamics of highly nonlinear birefringent AS$_2$S$_3$ CPCFC by employing POM. We have proposed three novel CPCFC designs by changing the air hole diameter along the slow axis for our investigation.  Numerical simulations of novel CPCFC designs predicted that, it is possible to reduce the coupling length and increase the birefringence by reducing the air hole diameter along the slow axis at 1.55 $\mu$m. Moreover, it should be noted that further reduction in diameter of the air hole shifts the $\beta_2$ from negative to positive. Also, in this work, we have analyzed the dynamics of the coupling characteristics of highly nonlinear birefringent PCF coupler using POM in detail. From the pulse parameter equations derived using POM from governing CNLSE, the coupling characteristics are explored for two cases. Case: (i) At $\theta$ = $\pi/4$ the coupling characteristics of the individual pulse parameters are explored. It has been shown that the system behaves as tuned coupler with equal splitting ratio for both the polarization components. Moreover, all the designs exhibit good coupling characteristics. Our numerical calculation shows that the design with shorter coupling length and high birefringence exhibits the polarization splitting within shorter propagation length compared to the other designs.  Case: (ii) By varying $\theta$, we have demonstrated the possibility of controlling splitting ratio and provide a way to control the desired polarization component by introducing the input pulse with apt polarizing angle. In addition to the splitting ratio, the proposed structure also found to exhibit different amplitudes for different polarization components for $\theta$ variation. Thus, the projection operator technique serves as an ideal tool for describing the dynamics of coupling characteristics of PCF coupler for various angles and pump energies of the input pulse.

\section*{Acknowledgment}
KP acknowledges DST, IFCPAR, NBHM, DST-FCI and CSIR, the Government of India for financial support through major projects. RVJ Raja wishes to thank DST fast track programme for providing financial support.


\section*{References}
\begin{thebibliography}{10}
\bibitem{Agrawal} Agrawal G P 2013 \textit{Applications of nonlinear fiber optics} (NewYork: Academic Press)
\bibitem{Saitoh1} Saitoh K, Sato Y and Koshiba M 2003 \textit{Opt. Exp.} \textbf{11} 3188
\bibitem{Gerome} G\'{e}r\^{o}me F, Auguste J -L and Blondy J -M 2004 \textit{Opt. Lett.} \textbf{29} 2725
\bibitem{Fogli} Fogli F, Saccomandi L, Bassi P, Bellanca G and Trillo S 2002 \textit{Opt. Exp.} \textbf{10} 54
\bibitem{Zhang} Zhang L and Yang C 2003 \textit{Opt. Exp.} \textbf{11} 1015
\bibitem{Uthayakumar1} Uthayakumar T, Raja R V J and Porsezian K 2012 \textit{J. Lightwave Tech.} \textbf{30} 13
\bibitem{Uthayakumar2} Uthayakumar T, Raja R V J and Porsezian K 2013 \textit{Opt. Commun.} \textbf{296} 124
\bibitem{Saitoh2} Saitoh K, Florous N J, Varshney S K and Koshiba M 2008 \textit{J. Lightwave Tech.} \textbf{26} 663
\bibitem{Gerosa} Gerosa R M, Biazoli C R, Cordeiro C M B and de Matos C J S 2012 \textit{Opt. Exp.} \textbf{20} 28981
\bibitem{Ortigosa} Ortigosa-Blanch A, Knight J C, Wadsworth W J, Arriaga J, Mangan B J, Birks T A, and Russell P St J 2000 \textit{Opt. Lett.} \textbf{25} 1325
\bibitem{Hansen2} Hansen T P, Broeng J, Libori S E B, Knudsen E, Bjarklev A, Jensen J R, and
Simonsen H 2001 \textit{IEEE Phot. Tech. Lett.} \textbf{13} 588.
\bibitem{Ju} Ju J, Jin W, and Demokan M S 2003 \textit{IEEE Phot. Tech. Lett.} \textbf{15} 1375
\bibitem{Chen} Chen D and Wu G 2010 \textit{Appl. Opt.} \textbf{49} 1682
\bibitem{Liou} Liou J, Huang S and Yu C 2010 \textit{Opt. Commun.} \textbf{283} 971
\bibitem{Saitoh3} Saitoh K and Koshiba M 2002 \textit{IEEE J. Quan. Elec.} \textbf{38} 927
\bibitem{Anton} Husakou A V and Hermann J 2003 \textit{Appl. Phys. B} \textbf{77} 227
\bibitem{Monro} Monro T M, Richardson D. J. 2003 \textit{Comptes Rendus Physique} \textbf{4} 175
\bibitem{Dabas} Dabas B and Sinha R K 2010 \textit{Opt. Commun.} \textbf{283} 1331
\bibitem{Kanka} Ka\v{n}ka J 2008 \textit{Opt. Exp.} \textbf{16} 20395
\bibitem{Chremmos} Chremmos I D, Kakarantzas G and Uzunoglu N K 2005 \textit{Opt. Commun.} \textbf{251} 339
\bibitem{Uthayakumar3} Uthayakumar T, Raja R V J, Nithyanandan K and Porsezian K 2013 \textit{ICMAP2013} 1
\bibitem{Shuo} Shuo L, Shu-Guang L, Guo-Bing Y and Xiao-Yan W \textit{Chin. Phys. B} \textbf{21} 034217
\bibitem{Zakery} Zakery A and Elliot S R 2003 \textit{J. Non-Crys. Sol.} \textbf{330} 1
\bibitem{Li2} Li J H, Chiang K S and Chow K W 2014 \textit{Opt. Commun.} \textbf{318} 11.
\bibitem{Anderson} Anderson D 1983 \textit{Phys. Rev. A} \textbf{27} 3135
\bibitem{Kutz} Kutz J N, Holmes P, Evangelides S G, and Gordon J P 1998 \textit{J. Opt. Soc. Am. B} \textbf{15} 87.
\bibitem{Nakkeeran1} Wai P K A and Nakkeeran K 2004 \textit{Phys. Lett. A} \textbf{332} 239
\bibitem{Moubissi} Moubissi A B, Nakkeeran K, Dinda P T, and Kofane T C 2001 \textit{J. Phys. A: Math. Gen} \textbf{34}, 129.
\bibitem{dindacv} Kamagate A, Grelu P, Dinda P T, Soto-Crespo J M and Akhmediev N 2009 \textit{Phys. Rev. E} \textbf{79} 1.
\bibitem{Uthayakumar4} Uthayakumar T, Raja R V J, Nithyanandan K and Porsezian K 2013 \textit{Opt. Fiber Tech.} \textbf{19} 556
\bibitem{Li} Li J H, Chiang K S, Malomed B A and Chow K W 2012 \textit{J. Phys. B: At. Mol.  Opt. Phys.} \textbf{45} 165404
\end{thebibliography}
\end{document}